\documentclass[ba,preprint]{imsart}
\pubyear{2025}
\volume{TBA}
\issue{TBA}
\arxiv{2512.05940}
\firstpage{1}
\lastpage{1}

\RequirePackage[authoryear]{natbib}
\RequirePackage[colorlinks,citecolor=blue,urlcolor=blue,backref=page,backref=page]{hyperref}

\usepackage{amssymb,amsmath,amsthm}
\usepackage{graphicx}
\usepackage{svg}
\usepackage{lipsum}
\usepackage{bbm}
\usepackage{natbib}
\usepackage{subcaption}
\usepackage{siunitx}
\usepackage[most,skins]{tcolorbox}

\usepackage[capitalise]{cleveref}
\usepackage{booktabs}
\usepackage{wrapfig}

\newcommand{\iid}{\overset{\mathrm{iid}}{\sim}}
\newcommand{\V}[1]{\boldsymbol{\mathbf{#1}}}
\newcommand{\given}{\,|\,}
\newcommand{\klbar}{\, || \, }
\newcommand{\x}{ {\bf x} }

\newcommand{\Ytest}{ \V{Y}_{\text{test}} }
\newcommand{\Yn}{ \V{Y}_n }

\DeclareMathOperator{\trace}{tr}

\DeclareMathOperator{\KL}{KL}
\DeclareMathOperator{\cov}{cov}
\DeclareMathOperator*{\argmax}{arg\,max}

\DeclareMathOperator{\logdet}{log\,det}

\DeclareMathOperator{\EIG}{EIG}

\let\originalleft\left
\let\originalright\right
\renewcommand{\left}{\mathopen{}\mathclose\bgroup\originalleft}
\renewcommand{\right}{\aftergroup\egroup\originalright}

\startlocaldefs
\theoremstyle{plain}

\theoremstyle{definition}

\theoremstyle{remark}

\newtheorem{remark}{Remark}
\endlocaldefs

\begin{document}

\begin{frontmatter}

\title{Designing an Optimal Sensor Network via Minimizing Information Loss}
\runtitle{Designing Sensor Networks via Minimizing Info Loss}

\begin{aug}
\author{\fnms{Daniel} \snm{Waxman}\thanksref{addr1}\ead[label=e1]{daniel.waxman@stonybrook.edu}},
\author{\fnms{Fernando} \snm{Llorente}\thanksref{addr2}\ead[label=e2]{fllorente@bnl.gov}},
\author{\fnms{Katia} \snm{Lamer}\thanksref{addr3}\ead[label=e3]{klamer@bnl.gov}},
\and
\author{\fnms{Petar M.} \snm{Djuri\'c}\thanksref{addr1}\ead[label=e4]{petar.djuric@stonybrook.edu}}

\runauthor{D. Waxman et al.}

\address[addr1]{Department of Electrical and Computer Engineering, Stony Brook University, Stony Brook, NY 11790 USA
    \printead{e1} %
    \printead*{e4}
}

\address[addr2]{Computing and Data Sciences Directorate, Brookhaven National Laboratory, Upton, NY 11973 USA
    \printead{e2}
}

\address[addr3]{Department of Environmental and Climate Sciences, Brookhaven National Laboratory, Upton, NY 11973 USA
    \printead{e3}
}

\end{aug}

\begin{abstract}
Optimal experimental design is a classic topic in statistics, with many well-studied problems, applications, and solutions. The design problem we study is the placement of sensors to monitor spatiotemporal processes, explicitly accounting for the temporal dimension in our modeling and optimization. We observe that recent advancements in computational sciences often yield large datasets based on physics-based simulations, which are rarely leveraged in experimental design. We introduce a novel model-based sensor placement criterion, along with a highly-efficient optimization algorithm, which integrates physics-based simulations and Bayesian experimental design principles to identify sensor networks that ``minimize information loss'' from simulated data. Our technique relies on sparse variational inference and (separable) Gauss-Markov priors, and thus may adapt many techniques from Bayesian experimental design. We validate our method through a case study monitoring air temperature in Phoenix, Arizona, using state-of-the-art physics-based simulations. Our results show our framework to be superior to random or quasi-random sampling, particularly with a limited number of sensors. We conclude by discussing practical considerations and implications of our framework, including more complex modeling tools and real-world deployments.
\end{abstract}



\end{frontmatter}

\section{Introduction}

Problems in optimal experimental design (OED) have been classically studied in frequentist and Bayesian paradigms, resulting in a mature theory with hundreds of practical applications. However, these approaches typically rely on access to the true data or make model-free decisions. In parallel to the statistical development of experimental design, the proliferation of vast computational resources has profoundly impacted statistics and the sciences alike. For Bayesian statistics in particular, computational techniques such as Markov chain Monte Carlo and variational inference have led to a more widespread adoption of the Bayesian viewpoint within the broader statistics community \citep{martin2020computing}.  Increased computational resources have also spurred the burgeoning field of Bayesian deep learning \citep{papamarkou2024position}. In the sciences, complex, large-domain, physics-based numerical simulations represent one innovation unlocked by computational power -- for instance, in the climate sciences with the development of coupled earth system models \citep{golaz2019doe} and global cloud-resolving models \citep{satoh2019global}, and in protein folding research with the development of simulation models of all-atom molecular dynamics  \citep{lane2013milliseconds}.

The interplay between these two types of simulations --- statistical and physics-based --- is important in practice, as their capabilities are often complementary. Asymmetry in capabilities is often attributable to interpolation or extrapolation from data observed by sensors; whereas statistical techniques excel at interpolating observations, physical simulations are much better suited to extrapolating outside of a limited observation dataset, as their predictions are carefully constrained by scientific knowledge.\footnote{We note that while statistical models and physical simulations are not mutually exclusive, incorporating probabilistic estimation in physical simulations may be scientifically and/or computationally infeasible. Thus, for our purposes, we treat them as effectively separate objects.} Returning to the climate sciences, statistical approaches are well-suited to smoothing spatiotemporal observations, but generic approaches tend to produce high degrees of uncertainty at longer forecasting timescales \citep{pinder2023developments}.

In this work, we use data from physics-based models to determine the optimal locations for observing a spatiotemporal process using sensors. We assume that the physics-based models in question are, in some way, ``accurate enough'' for this task (an assumption that we formalize later), but are, in other ways, ``too inaccurate,'' thus driving a need to evaluate them against observations taken by sensors. To find the optimal locations for sensors, we will utilize recent advances in the variational inference of spatiotemporal Gaussian processes (GPs), which allow for efficient variational learning when spatial inducing points are fixed with respect to time and the spatiotemporal kernel is separable (i.e., decomposes as a product of a temporal and spatial kernel). The use of variational inference allows us to motivate the proposed method from an information-theoretic perspective in terms of ``information loss,'' while the use of physics-based simulation data can be interpreted as incorporating an ``inductive bias'' into our model.

This inductive bias is particularly relevant, as any model-based approach to OED necessarily relies on the quality of the model. Using physics-based simulations and variational inference allows us to create statistical models that maximally reconstruct posterior distributions under the best current knowledge available, rather than a more ignorant prior.

At this point, it is prudent to contextualize our method within the more general framework of Bayesian experimental design (BED). More precisely, BED deals with finding an ``optimal experimental design,'' which may include selecting the locations where a spatiotemporal process is observed, guided by information-theoretic criteria \citep{chaloner1995Bayesian}. However, BED typically targets intractable quantities, which makes optimization extremely difficult. In fact, some of the most popular utilities used in BED (e.g., the expected information gain) are doubly intractable, meaning they involve nested expectations over intractable posteriors. Therefore, even standard Monte Carlo methods cannot be applied in general \citep{rainforth2024modern}. 

As a result, much of the effort in BED is related to computational approaches, such as improved nested Monte Carlo estimators \citep{rainforth2018nesting} or approximate, amortized inference schemes \citep{foster2019variational}. An extremely common approximation is to optimize in a greedy, sequential manner, in which case BED is known as Bayesian adaptive design (BAD). However, in the case of selecting locations to observe spatiotemporal processes, this greedy selection may lead to undesirable behaviors.  

The crux of our argument is that, if accurate physics-based simulation data are available, they should inform experimental design. Such data are not typically incorporated into BED/BAD. We show that, fortunately, this alleviates computational burdens intrinsic to BED, resulting in tractable loss functions that are information-theoretically motivated. Additionally, our method can be used in combination with BAD as an informed ``starting point'' for the greedy optimization of arbitrary utility functions, as opposed to the common practice of using quasi-random samples.

This paper is structured as follows: in \cref{sec:surrogate_model_experimental_design}, we provide background on BED using surrogate models and particularly motivate our notion of optimality. In \cref{sec:ed_ours}, we introduce our ``minimum information loss (MIL)'' optimality criterion and discuss the properties of our method. In \cref{sec:spatiotemporal_gps}, we present a brief review of Gauss-Markov regression, focusing in particular on the recently proposed spatiotemporal variational Gaussian process (STVGP) \citep{hamelijnck2021spatio}, and show how it enables efficient computation of MIL-optimal solutions in spatiotemporal settings. We discuss more intimate connections with ``modern'' OED methods, including other variational ones.
Finally, we show the performance of our method on 
a practical case study in air temperature modeling in \cref{sec:case_study}, and conclude with a discussion and directions for future work in \cref{sec:discussion}. 
We also derive several novel methods for continuous sensor placement as baselines using STVGPs, as explicated in Appendix C \citep{waxman2025milsupplement}.

\section{Model-Based Experimental Design} \label{sec:surrogate_model_experimental_design}

In this section, we overview some existing methods for OED --- with a particular emphasis on Bayesian techniques --- and use this to motivate our proposed method. For simplicity, we first focus on the case where some real, scalar process $Y(\V{x}) \in \mathbb{R}$ is observed, indexed by some quantity $\mathbf{x}$ that belongs to a compact subset $\mathcal{X} \subset \mathbb{R}^{d_X}$. We outline existing methods related to the expected information gain (\cref{sec:ed_existing}), followed by a more general utility-based experimental design (\cref{sec:ed_utility}). An expert in Bayesian experimental design may consider skipping to \cref{sec:ed_ours} after familiarizing themselves with our notation.

\subsection{Existing Approaches in Bayesian Experimental Design} \label{sec:ed_existing}

In general, Bayesian experimental design aims to find the design $\xi\in\Xi$ that allows us to collect data $D\in\mathcal{D}$ and reduce our uncertainty about the parameter of interest $\theta \in \Theta$, through the use of a utility function $U(\xi,D,\theta)$ and a posterior distribution $p(\theta|D,\xi).$\footnote{Since these methods derive their optimality through a statistical model, we will call them ``model-based'' (juxtaposed to model-free approaches such as Latin Hypercube sampling).}
For learning about the process $Y(\x)$, the design problem refers to the task of selecting a suitable set of $n$ inputs $\V{X}_n \in \mathcal{X}^n$ where a process ${Y}(\cdot)$ is observed, resulting in observations $\Yn$, %
such that these observations are as informative as possible about the process, especially regarding unobserved outputs $\Ytest \triangleq \mathbf{Y}(\mathbf{X}_{\text{test}})$. In frequentist analysis, the notion of ``information'' is typically related to Fisher information \citep{pukelsheim2006optimal,box1982choice}, while Bayesian approaches center on Shannon's information theory instead \citep{chaloner1995Bayesian,rainforth2024modern}. 

In a Bayesian context, the most common criterion for optimality is differential entropy or (negative) information. Recall the differential cross-entropy of a random variable $A$ with pdf $p_A(x)$ and another random variable $B$ with pdf $p_B(x)$.\footnote{We will generally assume that $A$ and $B$ have densities with respect to the Lebesgue measure and are mutually dominating measures.} is given by
\begin{equation}
    H(A, B) \triangleq \mathbb{E}_A[-\log p_B(x)] = -\int_{\mathcal{A}} \left[\log p_B(x)\right] p_A(x) \, dx.
\end{equation}
If $A = B$, then the differential cross-entropy $H(A, B)$ is known simply as the differential entropy of $A$, denoted $H(A)$. Information-theoretically, the differential cross-entropy $H(A, B)$ accounts for the entropy of $B$ in coding for $A$. In other words, the differential cross-entropy gives the expected amount of information needed to encode samples from the true distribution, $p_A(x)$, using a code optimized for the distribution $p_B(x)$. The Kullback-Leibler (KL) divergence $\KL(p(A) \klbar p(B))$ relates the ``excess'' entropy in $H(A, B)$ with respect to $H(B)$:
\begin{align}
    \KL(p(A) \klbar p(B)) &\triangleq \mathbb{E}_A \left[ \log \frac{p_A(x)}{p_B(x)} \right] = H(A, B) - H(A).
\end{align}

The conditional (differential) entropy of a $B$ given $A$ is the expected entropy of the conditional distribution $p(b \given a)$, i.e., $
    H(B \given A) = -\mathbb{E}_{A, B} \left[\log p(b \given a) \right],$
and it provides a measure of the uncertainty (or randomness) remaining in $B$ after observing $A$.
The conditional entropy is further related to the mutual information $I(A; B) = H(B) - H(B \given A)$.
Similarly to the KL divergence, mutual information quantifies the amount of entropy of $B$ that is not explained conditionally by $A$.

The mutual information is particularly important in modern Bayesian experimental design, where the mutual information conditioned on the design parameters $\xi$ is optimized. The (conditional) mutual information is known as the expected information gain (EIG). In our case, where the design parameters are $\mathbf{X}_n$, the EIG is given by: 
\begin{equation} \label{eq:EIG_defn}
\EIG(\mathbf{X}_n) \triangleq \mathbb{E}_{\Yn \given \mathbf{X}_n}\left[H(\Ytest) - H(\Ytest \given \mathbf{Y}_n,\mathbf{X}_n)\right].
\end{equation}
This gain measures the expected reduction in uncertainty about a random variable ($\Ytest$) after observing new data ($\Yn$) that depend on $\mathbf{X}_n$.
The quantity %
$\EIG(\mathbf{X}_n)$ is generally difficult to compute since
\begin{equation}
    \EIG(\mathbf{X}_n) = %
    \mathbb{E}_{\Yn|{\bf X}_n} \left[ \mathbb{E}_{\Ytest|{\bf Y}_n, {\bf X}_n}\left[ \log p(\Ytest \given\Yn,{\bf X}_n) -  \log p(\Ytest) \right] \right]
\end{equation}
is a nested expectation with respect to the generally intractable marginal predictive distribution $p(\Ytest, {\bf Y}_n \given \mathbf{X}_n)$. It is thus a ``doubly-intractable quantity,'' which cannot be estimated with simple Monte Carlo methods \citep{rainforth2024modern}.

\subsection{Bayesian Adaptive Design}
To address the computational intractability of evaluating EIG, one often resorts to greedy optimization or simple statistical models. The greedy optimization approach, known as \emph{Bayesian adaptive design} (BAD), maximizes the EIG at index $n$ using previously selected sensor locations $\mathbf{X}_{n-1}$. This process identifies a new sensor location $\mathbf{x}_n$ and forms a new set of sensor locations $\mathbf{X}_{n} = \mathbf{X}_{n-1} \cup \{\mathbf{x}_n\}$, where location $\mathbf{x}_n$  is the location that maximizes $\EIG(\mathbf{X}_n)$.

If both \( \Ytest \) and the observations \( \Yn \) are jointly Gaussian, for example, under a GP prior with a Gaussian likelihood, then the expected information gain (EIG) simplifies considerably. In this case, the mutual information between \( \Ytest \) and \( \Yn \), conditioned on the design \( \mathbf{X}_n \), becomes
\begin{equation}
\label{eq:gaussian_eig}
    \EIG(\mathbf{X}_n) = \frac{1}{2} \left[ \log\det \mathbf{K}_{\text{prior}} - \log\det \mathbf{K}_{\text{posterior}} \right],
\end{equation}
where \( \mathbf{K}_{\text{prior}} = \cov(\Ytest) \) and \( \mathbf{K}_{\text{posterior}} = \cov(\Ytest \mid \Yn) \). These covariances depend on the selected design \( \mathbf{X}_n \), and this closed-form expression avoids the doubly intractable nested expectations.

If $\mathbf{X}_n$ and $\mathbf{X}_\text{test}$ are chosen to form a partition of $\mathcal{X}$, i.e., $\mathcal{X} = \mathbf{X}_n \sqcup \mathbf{X}_\text{test}$, then \cref{eq:gaussian_eig} becomes the mutual information $I(\mathbf{Y}_{\mathcal{X} \setminus \mathbf{X}_n} ; \Yn)$. While discrete optimization of this objective is still intractable, \citet{krause2008near} show that greedy optimization of the mutual information provides near-optimal sampling locations, with $(1 - 1/e)$ convergence. However, the greedy algorithm is \emph{not} guaranteed to converge if the choice in $\mathbf{X}_\text{test}$ is independent of the choice in $\mathbf{X}_n$, i.e., if the set of test locations is fixed and does not depend on the sensor network.

\begin{remark}
    The challenge in providing a tractable algorithm in \citet{krause2008near} is that the optimization problem is discrete, leading to combinatorial complexity. %
    To select $k$ sensor locations  out of  $n$ candidates, the greedy algorithm under a full GP model incurs a total computational cost of $\mathcal{O}(k n^4) $. This cost arises from two sources: at each of the $k$ greedy selection steps, the algorithm must evaluate the utility (mutual information) for up to $ n $ remaining candidates, which results in $\mathcal{O}(kn)$ utility evaluations. Each utility computation, in turn, generally requires $\mathcal{O}(n^3)$ time due to matrix inversions or log-determinant computations associated with GP inference. Thus, the total runtime scales as $\mathcal{O}(k n^4)$ when performed naively with full GP models. Even in the most generous case, their algorithm must explore $\mathcal{O}(k n)$ combinations of locations, which may still prove prohibitively expensive. On the contrary, our proposed approach is continuous in nature and therefore does not require a discretization of $\mathcal{X}$ for optimization.
\end{remark}

\subsection{Utility-Based Optimal Experimental Design} \label{sec:ed_utility}

Taking a more global view, model-based design criteria may maximize the expected value of some utility other than the information gain. Many can be interpreted under the umbrella of Bayes-optimal design, where the optimal observations maximize a utility function $U(\cdot)$ \citep{chaloner1995Bayesian},
\begin{equation}
    \mathbf{X}^\text{opt}_n = \argmax_{\mathbf{X}_n} \mathbb{E}_{\Yn, \Ytest} \left[ U(\Ytest, \mathbf{X}_n, \Yn)  \given \mathbf{X}_n \right].
\end{equation}
For example, we may rewrite the EIG in terms of the KL divergence between the posterior and prior of $\mathbf{Y}_\text{test}$ \citep{cover1999elements},
\begin{equation}
    \EIG({\bf X}_n) =
    \mathbb{E}_{\mathbf{Y}_n | \mathbf{X}_n}\left[ \KL(p(\mathbf{Y}_\text{test} \given \mathbf{Y}_n, \mathbf{X}_n) \klbar p(\mathbf{Y}_\text{test})) \right].
\end{equation}
Hence, the EIG uses the utility function 
\begin{equation}
    U_{\text{EIG}}(\mathbf{X}_n, \Yn) = \KL\left(p(\mathbf{Y}_\text{test} \given \mathbf{Y}_n, \mathbf{X}_n) \klbar p(\mathbf{Y}_\text{test})\right).
\end{equation}
In fact, this was the original motivation for EIG, pioneered by \citet{lindley1956measure}.

There are many other utility-based approaches that may be based on conditional entropy, posterior predictive variance, and more. Several of these criteria amount to placing different functional norms on the variance; for example, the \emph{integrated mean-square error} criterion minimizes the posterior predictive variance integrated over $\mathcal{X}$, and the \emph{maximum mean-square error} (MMSE) criterion minimizes the maximum posterior predictive variance \citep{sacks1989}. In either case, finding all the optimal sensor locations ($\mathbf{X}^\text{opt}_n$) in ``one shot'' is generally difficult, and greedy sequential approaches are used instead \citep[Ch. 6]{gramacy2020surrogates}. We discuss some additional utility-based Bayesian OED methods, most notably \emph{maximum entropy sampling} (MES) in Appendix A \citep{waxman2025milsupplement}

\section{Minimizing Information Loss: An Alternative Model-Based Utility} \label{sec:ed_ours}
In this section, we introduce our new optimality criterion, which ``minimizes information loss'' with respect to a simulator. We begin by motivating and mathematically defining the minimizing information loss (MIL) criterion (\cref{sec:motiv_and_def}) and its relationship to sparse GP regression (\cref{subsec:connect_to_svgp}), followed by a discussion of its core statistical assumptions (\cref{sec:assumptions}). Next, we present  two practical modalities of application that are particularly simple under the MIL framework: incorporating existing sensors and sensor removal (\cref{sec:ours_existing}). We follow in the next section (\cref{sec:spatiotemporal_gps}) by extending the MIL method to spatiotemporal design.

\subsection{Motivating and Defining the MIL Criterion} \label{sec:motiv_and_def}
The model-based design criteria discussed in \cref{sec:ed_existing} and their sequential implementations have several drawbacks that we attempt to address. One notable drawback of the aforementioned design criteria (and MES, in particular) is that they
tend to resemble na\"ive space-filling designs that do not depend on the data or statistical models \citep[Ch. 4]{gramacy2020surrogates}. Intuitively, this can be undesirable if the observed process $Y(\cdot)$ is spatially nonstationary, with some areas being more difficult to predict than others.
Indeed, in their work with GPs, \citet{krause2008near} show empirically that using nonstationary kernels while maximizing mutual information significantly improves performance and argue theoretically that stationary kernels result in space-filling designs. 

Furthermore, when using any Bayesian model, we typically have a set of hyperparameters $\boldsymbol{\psi}$ that must be optimized. Utility-based approaches generally do not address the fitting of hyperparameters and rely on a pre-specified model. We propose an approach in which the model hyperparameters, $\boldsymbol{\psi}$, are jointly learned along with the optimal sensor locations $\mathbf{X}_n$.

We work in a setting where simulations of an observable process, $\widetilde{{Y}}(\cdot)$, are available from a physics-based model. For now, let us make the (strong) assumption that the simulations are a good representation of the true observations ${Y}(\cdot)$ so that we may obtain an approximation $\widehat{{Y}}(\cdot)$ using only the simulations, which are more abundant than observations (for example, those collected in the field or the laboratory). In other words, we assume that the posterior distribution after observing %
input-output pairs $(\mathbf{X}_s, \mathbf{Y}_s)$ is well-approximated by the posterior over simulated data 
$(\mathbf{X}_s, \widetilde{\mathbf{Y}}_s)$, i.e., when evaluated pointwise over $\mathcal{X}$,
\begin{equation} \label{eq:posterior_approx}
    p(Y(\mathbf{x}) \given \mathbf{Y}_s, \mathbf{X}_s) %
    \approx p(Y(\mathbf{x}) \given 
    \widetilde{\mathbf{Y}}_s, \mathbf{X}_s).
\end{equation}
Then, we can select the design that is optimal in order to ``sparsify'' the complete statistical model, obtaining a set of optimal observation locations $\mathbf{X}_n^\text{opt}$ such that $|\mathbf{X}_n^\text{opt}| \ll | \mathbf{X}_s|$, where $|{\bf X}|$ represents the cardinality of the set of observation locations. In particular, we will work {\em backwards} in formulating a criterion that minimizes the loss of information when going from the full (simulation) posterior to the sparsified (simulation) posterior.

At this point, the reader may wonder what the point of obtaining additional observations is if \cref{eq:posterior_approx} already holds. The key to our approach is that the optimal locations are obtained through a sparsification of $\mathbf{X}_s$, which will be motivated information-theoretically shortly. As such, it is not necessary, in the end, for \cref{eq:posterior_approx} to hold directly. Rather, the working assumption is that the true posterior $p(Y(\mathbf{x}) \given \mathbf{Y}_s, \mathbf{X}_s)$ can be well-approximated with the same set of observation locations as the simulated %
$p(Y(\mathbf{x}) \given \widetilde{\mathbf{Y}}_s, \mathbf{X}_s)$,
which is a more granular property. 

The approximation in \cref{eq:posterior_approx} suggests the following utility, where the KL divergence between the posterior observing all $\mathbf{X}_s$ points and the posterior after observing only $\mathbf{X}_n$ is minimized,
\begin{equation} \label{eq:our_util}
    U(\mathbf{X}_n, \Yn) = 
    - \KL\left( p(\mathbf{Y}_\text{test} \given \widetilde{\mathbf{Y}}_n, \mathbf{X}_n) \klbar p({\mathbf{Y}}_\text{test} \given \widetilde{\mathbf{Y}}_s, \mathbf{X}_s) \right).
\end{equation}
Our method relies on the resemblance of \cref{eq:our_util} to a variational approximation used in the study of GPs.

\subsection{Connection to Sparse Variational Gaussian Processes} \label{subsec:connect_to_svgp}
To motivate this connection, we first recall the setting of sparse GP regression. Consider the scenario where we would like to fit a GP $f(\cdot)$ on a dataset $\mathcal{D} = \{(\mathbf{x}_n, y_n)\}_{n=1}^N$, where $N$ is very large. We will use $\mathbf{X}_N$ to refer to the inputs of $\mathcal{D}$, and $\mathbf{Y}_N$ to refer to the outputs. In this case, exact GP interpolation may be prohibitively expensive, as the computational complexity of posterior predictions is $\mathcal{O}(N^3)$ \citep{rasmussen_and_williams}. To simplify computations, we introduce a set of $M \ll N$ \emph{inducing points} $\mathbf{Z}_M = \{ {\bf z}_m \}_{m=1}^M \subset \mathcal{X}$. Denoting the function values at $\mathbf{X}_N$ as $\mathbf{f}_N$ and the function values at $\mathbf{Z}_M$ as $\mathbf{u}_M$, the posterior predictive of the full GP is then approximated as 
\begin{equation}
    p(y_* \given {\bf Y}_N,{\bf X}_N,\x_*) 
    \approx 
    \int p(y_* \given {\bf u}_M, {\bf Z}_M,\x_*) p({\bf u}_M \given {\bf Y}_N,{\bf X}_N, {\bf Z}_M)\, d{\bf u}_M.
\end{equation}

We note two things in the above approximation: first, the values $\mathbf{u}_M$ are treated probabilistically, and second, the predictions depend only on the $M$ inducing points $\mathbf{Z}_M$ rather than the much larger $\mathbf{X}_N$. Contrary to MES, however, we still incorporate the observed data in $p({\bf u}_M \given {\bf Y}_N,$ ${\bf X}_N, {\bf Z}_M)$. One method to determine inducing points is by optimizing the objective,
\begin{align} \label{eq:free_energy_gp}
    \mathcal{L}(\mathbf{Z}_M) &= \log \mathcal{N}\left(\mathbf{Y}_N \given \mathbf{0}, \sigma^2 \mathbf{I} + \widetilde{\mathbf{K}}_N\right) \notag \\
    &\stackrel{+}{=} -\mathbf{Y}_N \left(\widetilde{\mathbf{K}}_N + \sigma^2_\text{obs} \mathbf{I}\right)^{-1} - \logdet \widetilde{\mathbf{K}}_N.
\end{align}
where $\stackrel{+}{=}$ denotes equality up to a constant factor, and $\widetilde{\mathbf{K}}_N$ is an approximation of the full kernel matrix $\mathbf{K}_N$ built using $\mathbf{Z}_M$. When $\widetilde{\mathbf{K}}_N = \mathbf{K}_N$, this expression reduces to the usual GP marginal likelihood. Different sparse algorithms consider different approximations of $\mathbf{K}_N$ \citep{quinonero2005unifying}, with a popular example being the Nystr\"om approximation. 

We observe a superficial resemblance between optimizing the objective of \cref{eq:free_energy_gp} and the various model-based utilities discussed in \cref{sec:ed_existing}: \cref{eq:free_energy_gp} consists of a regularization term based on the prior covariance matrix and a model fit term that encourages agreement between $\mathbf{Y}_N$ and the prior predictive mean. Nevertheless, we focus more specifically on the sparse \emph{variational} approach \citep{titsias2009variational}, which is known to guard against overfitting and has become the standard framework for sparse GPs.

The variational sparse GP approach approximates the full GP with a variational objective, seeking the best approximation to the joint posterior 
\begin{equation} \label{eq:full_posteropr}
    p(\mathbf{f}_N, \mathbf{u}_M \given \mathbf{Y}_N, \mathbf{X}_N, \mathbf{Z}_M) = p(\mathbf{f}_N \given \mathbf{u}_M, \mathbf{Z}_M,  \mathbf{Y}_N, \mathbf{X}_N) p(\mathbf{u}_M \given \mathbf{Y}_N, \mathbf{Z}_M).
\end{equation}
If we assume $\mathbf{u}_M$ to be a sufficient statistic of $\mathbf{f}_N$, the joint posterior distribution can be factorized exactly as 
\begin{equation}
    p(\mathbf{f}_N, \mathbf{u}_M \given \mathbf{Y}_N, \mathbf{X}_N, \mathbf{Z}_M) = p(\mathbf{f}_N \given \mathbf{u}_M, \mathbf{X}_N, \mathbf{Z}_M) p(\mathbf{u}_M \given \mathbf{Y}_N, \mathbf{Z}_M).
\end{equation}

However, it is unlikely that $\mathbf{u}_M$ is a sufficient statistic in practice, and it holds only approximately. We, therefore, find a variational posterior distribution
\begin{align}
    q(\mathbf{f}_N, \mathbf{u}_M \given \mathbf{Y}_N, \mathbf{X}_N, \mathbf{Z}_M) &= p(\mathbf{f}_N \given \mathbf{u}_M, \mathbf{X}_N, \mathbf{Z}_M) q(\mathbf{u}_M \given \mathbf{Y}_N, \mathbf{Z}_M),
\end{align}
which is closest to the true posterior \eqref{eq:full_posteropr} in terms of the KL divergence. This variational distribution is obtained by solving
\begin{equation}
    \min_{q, \mathbf{Z}_M} \KL\left(q(\mathbf{f}_N, \mathbf{u}_M \given \mathbf{Y}_N, \mathbf{X}_N, \mathbf{Z}_M) \klbar p(\mathbf{f}_N, \mathbf{u}_M \given \mathbf{Y}_N, \mathbf{X}_N, \mathbf{Z}_M) \right).
\end{equation}

In the case of Gaussian likelihoods with variance $\sigma_{\text{obs}}^2$, this minimization problem is equivalent to maximizing the evidence lower bound (ELBO),
\begin{align}
    \mathcal{L}_V\left({\bf Z}_M\right)
    &=
    \log \left[\mathcal{N}\left({\bf Y}_N \mid \mathbf{0}, \sigma_{\text{obs}}^2 \mathbf{I}+\mathbf{Q}_{N N}\right)\right]-\frac{1}{2 \sigma_{\text{obs}}^2} \operatorname{Tr}(\mathbf{K}_\text{post}({\bf X}_N)), \nonumber \\
    & \overset{+}{=} 
    -\frac{1}{2}{\bf Y}_N^\top\left( \mathbf{Q}_{NN} + \sigma^2_\text{obs} \mathbf{I} \right)^{-1}{\bf Y}_N 
    -\frac{1}{2} \logdet \left(\mathbf{Q}_{NN} + \sigma^2_\text{obs} \mathbf{I} \right)\notag\\
    &\hspace{0.4cm} - \frac{1}{2 \sigma_{\text{obs}}^2} \operatorname{Tr}(\mathbf{K}_\text{post}({\bf X}_N)), \label{eq:gp_vfe}
\end{align}
where $\mathbf{Q}_{NN}$ denotes the Nystr\"om approximation, and $\mathbf{K}_\text{post}({\bf X}_N) = \text{Cov}({\bf f}_N|{\bf u}_M)$ denotes the posterior covariance matrix at ${\bf X}_N$ given ${\bf u}_M$,
\begin{align}\label{eq:cov_nystrom}
    \mathbf{Q}_{N N} &= \mathbf{K}_{N M} \mathbf{K}_{M M}^{-1} \mathbf{K}_{M N}, \\
    \mathbf{K}_\text{post}({\bf X}_N) &= \mathbf{K}_{NN}-\mathbf{K}_{NM} \mathbf{K}_{MM}^{-1} \mathbf{K}_{MN},\label{eq:cov_variational}
\end{align}
where $\mathbf{K}_{NN}$, $\mathbf{K}_{MM}$ are kernel matrices computed using ${\bf X}_N$ and ${\bf Z}_M$, respectively, and $\mathbf{K}_{MN} = \mathbf{K}_{NM}^\top$ is the rectangular matrix containing cross-covariances. Note that $\mathcal{L}_V(\mathbf{Z}_M)$ is only a function of ${\bf Z}_M$ since, in the case of a Gaussian variational family, the optimal expression for the variational distribution 
 is available in closed form, $q_\texttt{opt}({\bf u}_M|{\bf Y}_N, {\bf X}_N, {\bf Z}_M) = \mathcal{N}\left( {\bf u}_M| \boldsymbol{\mu}, \mathbf{A} \right)$, for known $\boldsymbol{\mu}$ and $\mathbf{A}$. 

Crucial for its use as a model-based utility function, the maximization of the ELBO in \cref{eq:gp_vfe} only involves $\mathbf{Y}_N$ through simple matrix algebra. This allows us to write the KL divergence term as a model-based utility for which the empirical risk is an unbiased Monte Carlo estimate. In particular, we treat $\mathbf{X}_N$ as a set of inducing point locations $\mathbf{Z}_M$ and use the utility,
\begin{equation}
    U_{\text{MIL}}({\bf Z}_M) = \min_q \KL\left(q(\mathbf{f}_N, \mathbf{u}_M \given \mathbf{Y}_N, \mathbf{X}_N, \mathbf{Z}_M) \klbar p(\mathbf{f}_N, \mathbf{u}_M \given \mathbf{Y}_N, \mathbf{X}_N, \mathbf{Z}_M) \right),
\end{equation}
where $({\bf X}_N,{\bf Y}_N)$ are observations derived from physics-based simulation data.

Once sensors have been deployed in the field at locations optimized using the simulated dataset, field observations may be used to generate predictions for the entire spatial domain (and not just the locations where they are installed) by retaining the covariance structure learned from the simulated dataset in $q(\mathbf{u})$, but replacing the mean values $\mathbb{E}[q(\mathbf{u})]$ with the observations. This approach is valuable as it effectively eliminates the need to perform computationally expensive physics-based models to map observations for the entire spatial domain, i.e., we obtain an efficient surrogate model as part of our optimization process. We note that this approach will only yield usable predictions within the time-frame in which we have measurements. In other words, while we obtain a convenient spatial interpolator, physics-based models are still the preferred approach for long-term climate projections.

We call this utility \emph{minimum information loss (MIL)}, relating it to the classical measure of EIG directly; while EIG seeks to maximize the expected information gain of the posterior with respect to the prior, MIL minimizes the information \emph{lost} between the full posterior and an approximate posterior taken (from the physics-based model or from observations) at the optimal sensor locations.

\subsection{``Accurate, But Not Accurate Enough'' Simulators} \label{sec:assumptions}

As noted in our derivations, the use of MIL-optimality requires a simulator that is  ``accurate enough'' for sensor placement design, but ``not accurate enough'' for full-scale scientific policy. We argue here that this situation is not uncommon in practice and formalize this notion for MIL-optimality.

In the end, we want to argue that the sparse variational formulation over a simulator distribution can be used for predictions on real data. We will require two approximations for this: first, a high signal-to-noise ratio (e.g., in the sense of the process variance to noise) in the real and simulated data, such that we can derive reasonable inducing point means from real observations. Second, we will require that the simulator is structurally accurate, such that errors are minor and spatially diffuse. To this end, we will undertake a perturbative analysis to show that, under small and diffuse simulator error, the corresponding ELBO does not change much.

Our first assumption regards signal-to-noise ratios, allowing us to replace the inducing point means $\V\mu = \mathbb{E}[q(\V{u})]$ with observations. Recall the analytic solution of $\V\mu$,
\begin{equation}
    \V\mu=\sigma^{-2} \V{K}_{MM} \left(\V{K}_{MM} + \sigma^{-2} \V{K}_{MN} \V{K}_{NM}\right)^{-1} \V{K}_{MN} \V{y},
\end{equation}
where $\sigma^2$ is determined as the estimated noise variance plus a regularizer $\trace{\V{K}_{\text{post}}} / N$. So long as the sparse variational approximation is good, we expect $\trace{\V{K}_{\text{post}}} / N$ to be small, and thus $\sigma^2$ to be close to the estimated noise level. Then, with a high signal-to-noise ratio in the simulation, an observation $y(\V{z}_m)$ will approximately be $\mu_m$. 

If we additionally assume a high signal-to-noise ratio in observations of the real process, we can thus approximate the corresponding ${\mu}_m$. This is a surprisingly realistic assumption in many settings: for example, in temperature observations, noise may be on the scale of a few tenths of a degree Celsius.

We have now shown that the mean of the variational posterior can be well-determined if the observations are taken from a high signal-to-noise ratio process. What remains is to argue that the variational covariance is appropriate.

We proceed with a perturbative analysis, investigating the effect of data perturbation $\delta_{\V{Y}} \triangleq \tilde{\V{Y}}_N - \V{Y}_N.$
In the collapsed bound \cref{eq:gp_vfe},  the data enter only through the data fit term \begin{align}
D(\mathbf{Y}_N) \triangleq \frac{1}{2} \V{Y}_N^\top (\V{Q}_{NN} + \sigma^2_\text{obs}\mathbf{I})^{-1} \V{Y}_N.
\end{align}
Thus, we investigate perturbations in the variational free energy (VFE) 
\begin{align}
\Delta \mathcal{L} \triangleq \mathcal{L}_{V}(\V{Z}_M; \V{Y}_N) - \mathcal{L}_V(\V{Z}_M; \V{Y}_N + \delta_{\V{Y}})
\end{align}
through perturbations of the data fit term. Since $\mathcal{L}_V$ depends on $\mathbf{Y}_N$ only through $-D(\mathbf{Y}_N)$, we have $\Delta\mathcal{L}=D(\mathbf{Y}_N+\delta_{\mathbf{Y}})-D(\mathbf{Y}_N)$; since $D(\cdot)$ is a quadratic form, 
\begin{align}
    \Delta \mathcal{L} %
    &=  \left[\nabla D(\V{Y}_N)\right]^\top \delta_{\V{Y}} + D(\delta_{\V{Y}})
\end{align}
where $\nabla D(\V{Y}_N) = (\V{Q}_{NN} + \sigma^2_\text{obs}\mathbf{I}
)^{-1} \V{Y}_N$.

This perturbation consists of the addition of two terms: 
\begin{enumerate}
    \item A linear term, $\left[\nabla D(\V{Y}_N)\right]^\top \delta_{\V{Y}}$. There are two notable ways for this term to be small. The first is that both $\nabla D(\V{Y}_N)$ and $\delta_{\V{Y}}$ are small, which can occur when the simulator is accurate %
    (hence $\delta_{\V{Y}}$ is small). Potentially more interesting is the case in which $\nabla D(\V{Y}_N)$ and $\delta_{\V{Y}}$ are nearly orthogonal under the metric induced by $(\mathbf{Q}_{NN} + \sigma^2_{\text{obs}}\mathbf{I})^{-1}$.
    This situation may be interpreted as the real data deviating from the simulator in directions that are well captured by the covariance structure.
    \item A quadratic term, $D(\delta_{\V{Y}})$. This term is small precisely when $\delta_{\V{Y}}$ has small energy under the inverse covariance $(\mathbf{Q}_{NN} + \sigma^2_{\text{obs}}\mathbf{I})^{-1}$, that is, when the perturbation lies primarily in directions of high posterior uncertainty. %
\end{enumerate}
In total, this indicates that the simulator error is structured in such a way that it it primarily lies in directions of high variance %
under the covariance $(\V{Q}_{NN} + \sigma^2_\text{obs}\mathbf{I})$. Equivalently, the error has low energy under the corresponding inverse covariance. Unfortunately, it is difficult to test this property directly, but it can be informed via domain expertise or empirical evaluations. This limitation is inherent to our ``pre-observation'' setting, as no reliable statistical test can be designed without data learned from subsequent deployment. Nonetheless, we provide the following heuristics to help understand when this assumption is plausible:

\begin{enumerate}
    \item {Predictive smoothness and stability:} The simulator should produce spatiotemporally coherent outputs that reflect realistic physical variability, with moderate gradients and low-frequency noise. Sharp, erratic, or discontinuous behavior may indicate poor generalization and reduce the reliability of the variational structure learned from simulation.

    \item {High signal-to-noise ratio:} The outputs of the simulator should dominate internal numerical noise or model uncertainty. This property can be checked empirically by computing the signal-to-noise ratios across simulated realizations or by estimating noise levels during GP training.

    \item {Non-localized errors:} The simulator may exhibit errors, but those should be spatially diffuse rather than concentrated in specific regions. Our perturbative analysis (Section~3.2) shows that, under small perturbations, non-localized errors tend to have little effect on the ELBO and the resulting inducing points.
    
    \item {Coherence with sparse observations:} If a limited number of real-world measurements are available, comparing them to simulator outputs can help establish empirical plausibility. In particular, modest correlation or bounded deviations support the assumption that simulation-informed placement remains meaningful.
\end{enumerate}

Note that small and non-localized errors may not be sufficient for a simulator to be useful in real-world decision making without the incorporation of real data. One potential reason for this is computational constraints: simulators in many scientific fields are often extremely expensive and their direct use may introduce prohibitive costs in decision time, even if they are highly accurate. In this case, the simulator may indeed be ``accurate enough,'' but not under realistic computational constraints. 

Another limitation arises when non-localized errors that are small in absolute terms but nonetheless influence decision-making. For example, a simulator may accurately capture the relative temperature between two areas (i.e., their covariance structure), yet fail to capture their absolute magnitudes, which is essential for downstream applications. 

Finally, we note that many other OED methods rely on assumptions about the reasonableness of the prior. In contrast, MIL optimality can be interpreted as finding a design that performs well under a prior-predictive check, effectively aligning the design with the simulator-informed prior up to the limits of current modeling knowledge.%

We include experiments investigating synthetic simulator error, reinforcing our theoretical analysis above, in Appendix E \citep{waxman2025milsupplement}.

\subsection{Practical Considerations: Existing Sensors \& Sensor Removal} \label{sec:ours_existing}

The MIL framework is flexible and amenable to several practical extensions to deal with existing sensors, count, or heavy-tailed data, and sensor removals.

In many practical applications, the search for an optimal large-scale experimental design may come only after some sensors have already been deployed at seemingly suboptimal locations; thus, when deciding on optimal locations to place new sensors, it may be valuable to consider the locations of existing sensors to leverage their observations. This is a sort of ``conditional'' optimality, which may be difficult to incorporate in other non-greedy approaches.

Fortunately, this issue is relatively simple to solve within our framework. The simple solution is to fix some subset of inducing points during optimization to the locations of pre-existing sensors. In this sense, we find the optimal variational approximation in which some subset of observation locations is known and fixed \emph{a priori}. Note that this does not result in additional computations, nor does it interfere with solving for all $\mathbf{X}_N$ simultaneously. From an implementation standpoint, this can be achieved by simply not propagating the gradients with respect to existing observation locations. This process is discussed and illustrated in Appendix B \citep{waxman2025milsupplement}.

\label{sec:our_method_existing_locations}

Additionally, we can utilize a heteroscedastic noise likelihood rather than the standard i.i.d. Gaussian to account for different sensor ratings. In particular, in many cases, new sensors might possess significantly greater observational accuracy (i.e., less noisy observations) than old sensors, which is also important in optimization.

Differing likelihoods may also be useful for non-Gaussian response data. For example, the sparse variational GP can also be used for other modalities, such as count data (e.g., with a Poisson likelihood) or with heavier tails (e.g., with a Student's-$t$ likelihood). This is another unique aspect of our work within the Bayesian OED literature.

One practical problem common to spatio(temporal) sensor placement is that of sensor removal: suppose it is too expensive to keep all $N$ placed sensors, and we must remove $R$ of them. The MIL-optimal condition is well-suited for this, as we can simply evaluate the optimization bound with each set of $N \choose R$ sensors removed, allowing us to remove sensors with an awareness of their downstream predictive effects.

\section{MIL-Optimal Sensor Placement for Spatiotemporal Processes}
\label{sec:spatiotemporal_gps}

To tackle the spatiotemporal modeling  requirements of our problem, we will use a recent variational spatiotemporal GP model proposed by \citet{hamelijnck2021spatio}. We first explain why the usual variational sparse GPs are insufficient for spatiotemporal settings (\cref{sec:ed_time_series}), before introducing sparse variational spatiotemporal GPs (\cref{sec:stsvgps}). We then discuss the implications for computing $U_\text{MIL}(\cdot)$.

\subsection{Initial Remarks About Spatiotemporal Processes}
 \label{sec:ed_time_series}
We now consider the setting where ${Y}(\mathbf{x})$ is a spatiotemporal series. To make this explicit in notation, we will separate $\mathbf{x}$ into the spatial component and temporal component, $(\mathbf{x}, t)$, and denote the observed process as ${Y}(\mathbf{x}, t)$.

In many scientific applications, it is desirable for $\mathbf{X}_N$ to refer to a set of \emph{spatially fixed} observation locations, repeatedly observed over time, for example, when designing a network of temperature sensors that record measurements hourly. While this is a seemingly innocuous requirement, it turns out to be quite limiting for existing methods.

For $U_\text{MIL}(\cdot)$ in particular, the usual optimization procedure becomes problematic, as the inducing points become irregularly distributed in space and time. Even after restricting to fixed spatial locations, computational problems arise, as the computation of the ELBO term in \cref{eq:gp_vfe} scales cubically with the number of inducing points $M$. Na\"ively, for a time series of length $N_t$, the sparse variational GP would be implemented with $M \propto N_t$, leading to potentially prohibitive $\mathcal{O}(N_t^3)$ computational complexity.

We thus formulate the following desiderata for $U_\text{MIL}(\cdot)$ in spatiotemporal settings:

\begin{tcolorbox}[enhanced,%
    colback=white,%
    colframe=black,%
    colbacktitle=white,%
    coltitle=black,%
    fonttitle=\bfseries, %
    title=Desiderata,%
    attach boxed title to top left={yshift=-2.5mm, xshift=0.6cm},
    boxed title style={
        frame hidden,       %
        boxrule=0pt        %
    }]
    \begin{enumerate}
    \item {(\textbf{Scalability})} Computations should scale well (e.g., linearly) with the length of the time series $N_t$. 
    
    \item {(\textbf{MIL-Optimality})} There should be a set of static \emph{spatial} inducing points $\mathbf{Z}_M$, repeated in time, which results in the best average variational bound across all time instances $t$.
    
    \item {(\textbf{Joint Optimization})} Optimization should be available in ``one-shot,'' i.e., the inducing points and hyperparameters should be optimized \emph{jointly}. 
\end{enumerate}
\end{tcolorbox}

These desiderata rule out existing approaches in multiple ways: for one, MIL-optimality is novel and requires fairly recent machinery from variational inference to optimize in a scalable fashion. Additionally, joint optimization has been, in general, difficult to achieve in BED-like approaches, leading to the widespread use of greedy algorithms like BAD or that of \citet{krause2008near}. In what follows, we introduce the necessary tools to satisfy our desiderata.

\subsection{Satisfying Our Desiderata} \label{sec:stsvgps}
A na\"ive application of variational GPs to the MIL objectives is quite unlikely to succeed; only recently have new GP methods been developed that can satisfy our desiderata, and the machinery remains nonstandard. Developing a computational framework for deriving MIL-optimal sensor configurations will consist of two steps: (1) spatiotemporal GP regression via stochastic differential equations \citep{sarkka2013spatiotemporal}, and (2) application of conjugate-computation variational inference (CVI) \citep{khan2017conjugate} with fixed inducing points \citep{hamelijnck2021spatio}. These two steps handle the scalability desiderata and the optimality desiderata, respectively. 

\paragraph*{Scalability}
Overcoming the cost of GP posterior inference --- which is $\mathcal{O}(N^3)$ as computed using the kernel --- is a big challenge in GP inference generally. This can be particularly problematic for long time series, where $N \propto N_t$ implies that datasets can grow quite quickly. For a special subset of kernels, however, it is possible to perform posterior predictive inference in linear (i.e., $\mathcal{O}(N_t)$) time.

This approach, popularized by \citet{sarkka2013spatiotemporal}, represents a GP in terms of a stochastic differential equation (SDE), where posterior inference reduces to Kalman filtering and smoothing. For the moment, we will consider a temporal GP $f$, indexed by a scalar $t$, which we assume to have a stationary kernel $\kappa(t, t') = \kappa(t - t')$. For many common kernels, we may form an augmented state $\overline{\mathbf{f}}$ (which often consists of the value of $f$ and some time derivatives of $f$) and a corresponding linear time-invariant SDE that describes the GP $f$. This SDE is usually marginalized to a discrete-time linear dynamical system, with linear observations,
\begin{align} \label{eq:markovian_gp_state_space}
    \overline{\mathbf{f}}(t_{n+1}) &= \mathbf{A}_n \overline{\mathbf{f}}(t_n) + \mathbf{q}_n,\\
    y_{t_n} &= \mathbf{H} \overline{\mathbf{f}}(t_{n}) + \varepsilon_{t_n},
\end{align}
where $\V{q}_n \iid \mathcal{N}(\mathbf{0}, \V{Q}_n)$ and $\varepsilon_{t_n} \iid \mathcal{N}(0, \sigma_n^2)$. The parameters of this linear dynamical system (i.e., the matrices $\mathbf{A}_n$ and $\mathbf{H}$, and the noise (co)variances $\mathbf{Q}_n$ and $\sigma_n^2$) are determined directly by the choice of kernel $\kappa(t - t')$ and the discretization interval.

GPs that can be written using \cref{eq:markovian_gp_state_space} are referred to as \emph{Markovian} GPs, or sometimes \emph{state-space} GPs. The latter term may be confused with state-space models in which GPs describe the state transition; thus, we will avoid it. Although not all GPs are Markovian, the state-space representation can be derived exactly for many standard choices of kernels -- for example, kernels from the Mat\'ern family -- and may be approximated for many others \citep{sarkka2013spatiotemporal}. 

After converting to this representation, \emph{exact} linear time inference can proceed via Kalman filtering and smoothing algorithms \citep{sarkka2023bayesian}. Furthermore, the marginal likelihood can be computed in the filtering step, so that the evaluation of likelihood-based optimization objectives also scales as $\mathcal{O}(N_t)$. In the spatiotemporal case, we can use infinite-dimensional filtering techniques, which also scale linearly with $N_t$ \citep{sarkka2013spatiotemporal}.

However, the use of Markovian GPs alone does not solve the rest of our desiderata. In particular, incorporating variational objectives that solve MIL-optimality is not straightforward within this framework. Fortunately, a recent application of CVI has shown this to be possible.

\paragraph*{MIL-Optimality and Joint Optimization} Despite their impressive $\mathcal{O}(N_t)$ scaling, spatiotemporal Markovian GPs still incur an $\mathcal{O}(N_s^3)$ computational cost, where $N_s$ is the number of spatial locations. This can be understood by recognizing that the state dimension scales linearly with $N_s$, and Kalman filtering/smoothing algorithms have cubic time complexity in their state dimension. This limitation and the presence of non-conjugate likelihoods have led to several works on sparse and variational approaches to Markovian GPs \citep{wilkinson2021sparse,chang2020fast}.

The main additional assumption we must make is that the GP kernel is separable: 
\begin{equation} \label{eq:separable_kernel}
    \kappa(\mathbf{x}, t, \mathbf{x}', t') = \kappa_s(\mathbf{x}, \mathbf{x}') \kappa_t(t, t').
\end{equation}
Probabilistically, this induces a tensor-product structure in the spatiotemporal covariance, while computationally it leads to state-space models
Kronecker-like structure. 

Using this assumption, \citet{hamelijnck2021spatio} apply CVI to obtain a spatiotemporal sparse variational GP (ST-SVGP) with favorable computational properties. CVI reformulates  variational inference as approximate Bayesian inference with respect to a conjugate exponential family model and allows efficient natural-gradient updates of the variational parameters.
In the case of ST-SVGPs, we fix the inducing point locations to a $M_s \times N_t$ spatiotemporal grid. The CVI algorithm then proceeds efficiently in linear-time along the temporal dimension (via Kalman filtering), 
followed by natural gradient updates of the pseudo-likelihood parameters. Remarkably, the resulting predictive distribution is exactly equivalent to that of a traditional VGP with inducing points on a spatiotemporal grid. 

With $\boldsymbol{\mu}_t$ and $\mathbf{A}_t$ denoting the variational mean and covariance of the inducing variables at time $t$, \cref{eq:cov_nystrom,eq:cov_variational} together with \cref{eq:separable_kernel} then imply 
\begin{equation}
    q(\mathbf{f}_t) = \mathcal{N}(\mathbf{m}_t, \boldsymbol{\Sigma}_t),
\end{equation}
where writing $\mathbf{K}_{NN}$, $\mathbf{K}_{NM}$, and $\mathbf{K}_{MM}$ for the appropriate kernel blocks at time $t$,
\begin{align}
    \mathbf{m}_t &= \mathbf{K}_{NM} \mathbf{K}_{MM}^{-1} \V{\mu}_t, \\
    \boldsymbol{\Sigma}_t &= \mathbf{K}_{NN}
    - \mathbf{K}_{NM}\mathbf{K}_{MM}^{-1}\mathbf{K}_{MN}
    + \mathbf{K}_{NM}\mathbf{K}_{MM}^{-1}\mathbf{A}_t\mathbf{K}_{MM}^{-1}\mathbf{K}_{MN}.
\end{align}

What remains, then, is obtaining $\mathbf{m}_t$ and $\mathbf{A}_t$ from the test data. It is clear that the value of $\mathbf{m}_t$ on the training data will typically be different from its value on the testing data. However, it is more realistic to assume that the covariance structure $\mathbf{A}_t$ remains intact. For all of our experiments, we reuse $\mathbf{A}_t$ directly and obtain $\mathbf{m}_t$ through a single pass of CVI on the test data, keeping $
\mathbf{A}_t$ fixed. Notably, different choices for determining $\mathbf{A}_t$ exist; for example, one may use additional simulation data that directly emulate the observed data.

\section{Comparisons to Existing Approaches}
Having introduced MIL-optimality, we now compare it to several approaches  for spatial and spatiotemporal OED, as described in the literature. We organize these into classical methods (\cref{sec:classical_compare}), other methods utilizing variational bounds (\cref{sec:variational_compare}), and various GP-based approaches (\cref{sec:others_compare}).

\subsection{Classical Approaches} \label{sec:classical_compare}

Using simulation to determine Bayesian optimal designs dates back to the seminal work of \citet{MULLER2005509}. An expected utility is approximated pointwise by averaging over samples drawn from the joint distribution of model parameters and observations, given the design. Then, a stochastic optimization algorithm or Monte Carlo sampling can be employed to approximate the solution of the optimization problem \citep{ryan2016review,HUAN2013288}. For some utility functions, such as EIG, we require nested Monte Carlo estimators for the pointwise evaluation of the criterion \citep{rainforth2018nesting}, which has motivated the use of {\it amortized} approaches \citep{rainforth2024modern}. In the GP context, popular utilities yield closed-form expressions such as MES or the discrete version of the IMSE \cite[Ch. 6]{gramacy2020surrogates}. However, optimizing these criteria can still be very challenging, having to resort to sequential and greedy approaches.  
{%
Related work in mobile sensor networks has employed sequential/adaptive sampling strategies based on classical criteria such as IMSE or MES \citep{xu2011sequential,nguyen2021mobile}.
}

Another potential difficulty is the need to select designs from a discrete set of possibilities.
\citet{krause2008near} proposes a greedy algorithm for maximizing the EIG over a discrete domain.
Recent works from the machine learning community focus on finding suitable acquisition functions (active learning) for GPs \citep{riis2022bayesian}.
In summary, in Bayesian OED (BOED) we require a model, previously fitted using some initial data, a computable utility/criterion, and a search algorithm. In classical BOED, these elements are usually realized independently, resulting in inefficient algorithms.

\subsection{Variational Approaches} \label{sec:variational_compare}

As discussed in \cref{sec:ed_existing}, the most popular criterion for BOED is the EIG. 
However, classical estimators of the EIG require a nested loop for each pointwise evaluation. These computational burdens have motivated a recent trend that aims at amortizing the search for the optimal design \citep{rainforth2024modern}. 
Leveraging variational approximations of the posterior or the marginal likelihood, \citet{foster2019variational,foster2020unified} derive variational bounds on the EIG that can be optimized over both variational parameters and designs, hence avoiding the need for nested estimators. More importantly, the computation and optimization of EIG are performed jointly.
 In this paper, we leverage simulation in a different way, as we do not assume real data are available to infer a reasonable model. Instead, we may view the MIL criterion as if simulation data arises from a black-box model of the real process, provided by the simulator, for which (hyper)parameters and the  utility are optimized jointly.

\subsection{Gaussian Process-Based Approaches} \label{sec:others_compare}

Though our application of sparse variational GPs using simulation data for spatiotemporal experimental design is, to our knowledge, novel, it has long been recognized that sparse GP learning appears similar to sensor placement tasks. For example, \citet{krause2008near} remarks that the original inducing point method of \citet{snelson2005sparse} is similar and more tractable. More recently, \citet{jakkala2023efficient} use the sparse variational lower bound, with no data, to geometrically determine inducing points based on the GP kernel.

One related approach using ST-SVGPs, in particular, was proposed by \citet{herrera2022spatial}, combining the ST-SVGP with the algorithm of \cite{krause2008near} to create a sensor network with high mutual information in an application to air-quality data. To do so, an ST-SVGP is first trained on a set of ``training points;'' Krause's algorithm is subsequently applied to a set of ``candidate points,'' and the ST-SVGP is retrained. 
To apply Krause's algorithm, the mean spatial covariance is used, i.e.,
\begin{equation}
    \bar{\V{\Sigma}} = \frac{1}{N_t} \sum \V{\Sigma}_t.
\end{equation}
\citet{herrera2022spatial} apply their method to a set of air quality sensors in London and find modest increases in the mutual information accordingly. 

While \citet{herrera2022spatial} is successful in creating sensor networks with high mutual information, their setting is fundamentally different from ours in that they assume the existence of real data (i.e., field observations) to train and optimize the networks. Additionally, the application of Krause's algorithm is not straightforward without identifying a set of discrete ``candidate'' locations, as computation grows cubically with the number of candidate points. In our experiments, we use baselines modified for continuous optimization accordingly. Furthermore, air temperature presents a different spatiotemporal pattern from air quality: the former shows a robust diurnal cycle, while the latter is largely influenced by point sources.

\section{Case Study: Air Temperature Modeling in Arizona} \label{sec:case_study}

The motivation for the MIL was to effectively utilize large amounts of physics-based simulation data in training a statistical model, which,in turn,is used to identify optimal sensor locations to observe a spatiotemporally varying process. 
One potential application for our proposed method lies in the climate sciences, where expensive physics-based simulations may be used to create a physically motivated statistical model that enables us to identify optimal sensor locations to observe air temperature, a spatiotemporally varying process. While these physics-based simulations are considered generally accurate, in the sense that they have been compared to historical observations and are physically realistic, they are often not thoroughly evaluated, both in space and time, and in terms of the several observables that they produce. As a result, it is valuable to collect observational data from new locations that are maximally informative in order to further evaluate and calibrate the physics-based model. Furthermore, physics-based simulations are often extremely computationally expensive, making it desirable to develop alternative methods to produce climate and weather maps by leveraging sparse but strategically positioned observations. This introduces the experimental design problem of identifying optimal sensor locations. All code used in our experiments is available under an MIT License at \url{https://github.com/DanWaxman/MILSensors}.

\subsection{The Physics-Based Model Data}

For this application, we rely on simulated data generated by the Weather Research Forecast (WRF) model \citep{powers2017weather}. WRF is a physics-based weather and climate model that has been used, among other things, to study heat waves and their interactions with the urban heat island (e.g., \citealp{li2014quality}) as well as the potential of green roofs as heat-mitigating solutions for improving urban resilience (e.g., \citealp{tewari2019interaction}).

\begin{wrapfigure}{r}{0.476\linewidth}
    \centering
    \includegraphics[width=\linewidth]{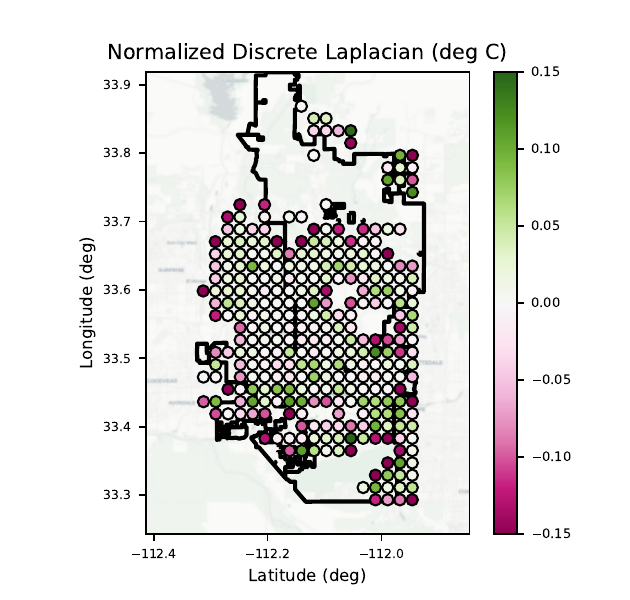}
    \caption{The degree-normalized discrete Laplacian of average temperature in the WRF-2013 dataset.}
    \label{fig:discrete_laplacian}
\end{wrapfigure}
Simulations were produced continuously from May 1st (00 UTC) to
August 31st (23 UTC) to capture extreme heat conditions and avoid irregular (in space and time) air temperature fluctuations that would result from storm activity typical of the North American monsoon season \citep{vera2006toward}. The month of May was considered a spin-up period and was discarded from subsequent analysis. The physics-based model domain was composed of three two-way nested domains centered on Arizona. Here, we focus on the urban grids of the inner domain, specifically the 314 grids encompassing Phoenix, Arizona. This inner domain has a spatial resolution of \SI{2}{\kilo\meter}. 
Additional information about the physics-based model setup and its parameterization can be found in \citet{salamanca2025effects} and \citet[Table 1]{salamanca2015summertime}, respectively.

To illustrate the degree of spatial nonstationarity in the 2-m air temperature of the city, we show the degree-normalized discrete Laplacian in \cref{fig:discrete_laplacian}. This shows that some areas are much more related to their neighbors than others. For example, locations near the river or at the edges of the city have higher degrees of spatial variation in air temperature and thus might be more difficult to predict. The task is to create an optimal sensor network for the observation of air temperature, controlling the predictive error.

Note that we did not assess the representativeness of the summer of 2013 for the climate of Arizona. As such, the results shown here may not be representative of a sensor network that would be optimal for observing air temperature in Phoenix for years and seasons other than those exhibiting the same climatic characteristics as the summer of 2013. The results shown here are merely a case study to demonstrate the applicability of our proposed approach. Future work will endeavor to include a broader simulation dataset to inform the placement of a sensor network that would provide comprehensive observations for a range of climate conditions, from dry years to wet years. 

\subsection{MIL-Optimal Points Provide Good Performance}

One main hypothesis of our work is that the inducing points in the variational approach will naturally gravitate towards areas that are difficult to predict and, therefore, provide reasonably informative observations, even in the prior kernel. To test this hypothesis, we find the MIL-optimal inducing point locations using data from June 1 to June 30, 2013, and subsequently treat the data from August 1 to August 31, 2013, as ground truth. We then compute the root-mean-squared error (RMSE) in air temperature and the negative predictive log-likelihood (NPLL). For both of these measures, lower values indicate better performance; intuitively, RMSE measures how good the mean performance is, and NPLL measures the performance of the predictive distribution.

\begin{remark}
    While we lack ``real data'' for comparison, the data from June 2013 and August 2013 are quite different, e.g., in terms of maximum daily temperatures, temperature range, etc. We thus argue that utilizing only June 2013 as ``simulator data'' (i.e., our training data) already imposes a minor effect similar to a simulation-to-real gap. 
\end{remark}

\begin{figure}
    \centering
    \includegraphics[width=0.9\linewidth]{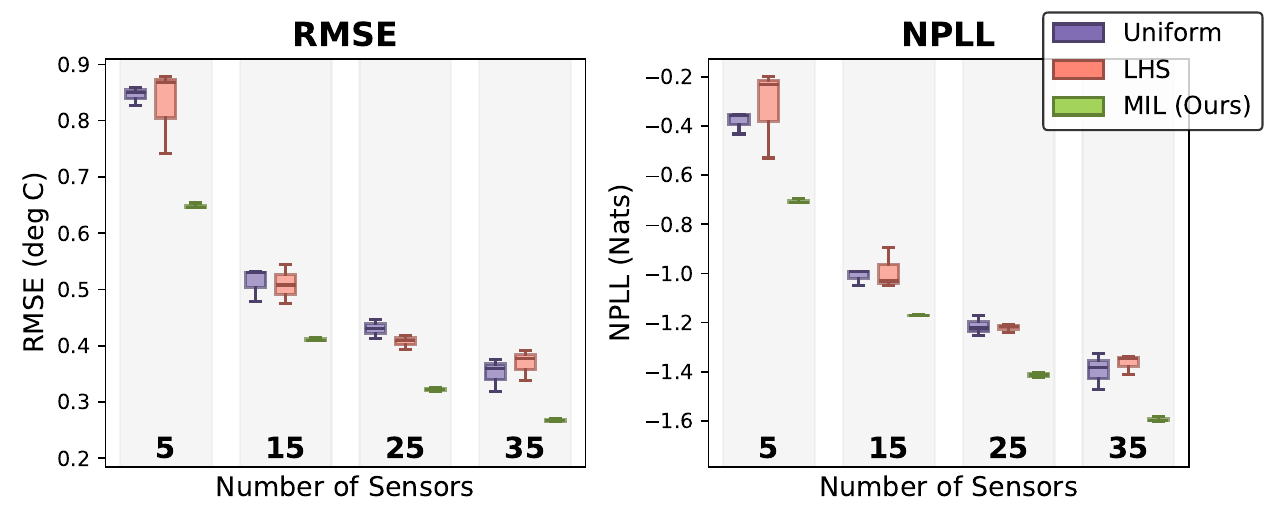}
    \caption{The RMSE and NPLL of uniform, LHS, and MIL-optimal strategies in our test problem with 5, 15, 25, and 35 sensor locations.}
    \label{fig:n_5_15_25_35_results}
\end{figure}

We compare our approach to four other alternative network design strategies. The first two are model-free approaches: a ``uniform'' strategy that selects observation points uniformly over the index set, and a Latin hypercube sampling (LHS) strategy, in which the LHS sample is taken, and the closest observation locations are selected. 

The other two are model-based approaches popular in spatiotemporal experimental design \citep{mateu2012spatio}, based on maximizing the predictive entropy of predictions at the design \citep{jin2012entropy}, and minimizing the predictive variance given the design \citep{heuvelink2012sampling}; we will refer to these methods as maximum entropy sampling (MES) and integrated minimum square error (IMSE) designs, respectively.
Applying these model-based strategies in a fair and consistent manner is not straightforward in our case, since our method has a fundamentally different setting from typical spatiotemporal OED. We discuss the choices we made and the computational improvements in Appendix C \citep{waxman2025milsupplement}.

Returning to MIL, we find that the learned test-time spatial covariance does not vary much over time, and we apply its values directly. Inducing point locations were initialized via k-means clustering using randomized initial points, which is the primary source of randomness in the MIL-optimal optimization procedure. At test time in the MIL-optimal strategy, we use the closest points in the dataset to each ``MIL-optimal'' point as ground truth.

For all comparisons, we use the kernel hyperparameters learned by the variational GP from the training dataset in the testing dataset. We use a mixture of kernels for the temporal dimension; a Mat\'ern-1/2 kernel for local and non-smooth changes in air temperature, and a Hida-Mat\'ern-3/2 kernel \citep{dowling2021hida} to help capture the diurnal cycle (i.e., daily quasiperiodicity) expected to be present in the data. In the spatial dimension, we use a Mat\'ern-3/2 kernel. We note that while the temporal dimension must be modeled by a Markovian GP, there is no such restriction on the spatial kernel. Indeed, we may even use nonstationary kernels such as the Gibbs kernel \citep{gibbs1998bayesian}, which has demonstrated success in modeling climate systems \citep{lachland2022kernel}. Jointly learning a nonstationary kernel and its inducing point representation introduces its own optimization challenges but does not fundamentally change the theoretical underpinnings of MIL-optimality. We therefore include some preliminary experiments using deep kernel learning \citep{wilson2016deep}, which learns nonstationary kernels using deep neural networks, in Appendix D \citep{waxman2025milsupplement}, but we leave a more detailed study as an interesting direction for future work.

We show the resulting RMSE and NPLL for three realizations of networks with 5, 15, 25, and 35 sensors, each beginning with different initialized locations (\cref{fig:n_5_15_25_35_results}). These results support our hypothesis that MIL-optimal inducing point locations provide strong performance relative to random or quasi-random sensor networks, particularly with a low number of sensor locations. This is true for both RMSE and NPLL. For example, using 15 sensors, the median RMSE drops from \SI{0.49}{\celsius} to \SI{0.41}{\celsius}.

\begin{figure}
    \centering
    \includegraphics[width=0.9\linewidth]{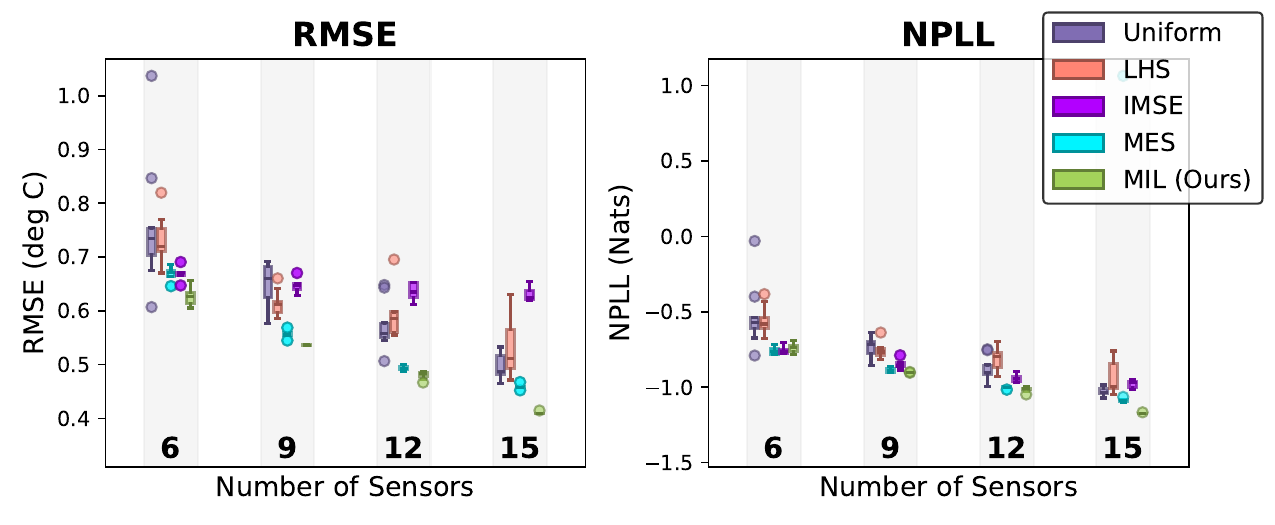}
    \caption{The RMSE and NPLL of uniform, LHS, IMSE, MES, and MIL-optimal strategies in our test problem with 3, 6, 9, 12, and 15 sensor locations.}
    \label{fig:n_3_6_9_12_15}
\end{figure}

Since in our initial experiment, the largest gains were made with smaller numbers of sensors, we repeated this experiment with 3, 6, 9, 12, and 15 sensors. Results are shown in \cref{fig:n_3_6_9_12_15} and are overall similar to the results of the previous experiment. In particular, the MIL-optimal configuration outperforms random configurations by a considerable margin on average, typically outperforming even the best of 10 uniformly random configurations. In this set of experiments, we additionally compare to MES and IMSE designs. We find that MES performs competitively with a small number of sensors (where space-filling designs are expected to work well); however, MIL becomes the dominant strategy by the time 15 sensors are considered. IMSE performs quite poorly in terms of RMSE, even compared to random and quasi-random design selections --- we believe this is due to optimization difficulties, which resulted in designs that do not tend to cover the corners of our domain. In any case, in terms of both RMSE and NPLL, MIL-optimal configurations possess a clear advantage by the time 15 sensors are placed.

Another feature of the MIL-optimal locations is their relatively small variance in performance, which suggests some convergence to optimal states. We visualize this in \cref{fig:maps_to_optimal}, where the initial states and their corresponding MIL-optimal states are shown. We see that inducing point locations often travel quite far from their initialization points. Interestingly, the random initialization point shown in purple results in a slightly different MIL-optimal location compared to the other two random initializations, but it yields comparable performance. 
This suggests that there may be multiple slightly different sensor network configurations that are locally MIL-optimal and perform similarly well. This is apparent in the extremely narrow distribution of RMSE between different initializations in \cref{fig:n_3_6_9_12_15,fig:n_5_15_25_35_results} when compared to (quasi-)random sensor networks. 

On account of the STVGP backbone, MIL-based optimization is extremely efficient. Computing solutions using $N_t \cdot N_s = 720 \cdot 314 \approx 2 \cdot 10^5$ data points and $9$ sensor locations takes approximately 5 minutes on modern GPU hardware (Nvidia A100). The computational complexity of the ST-SVGP is $\mathcal{O}(N_t M_s^2 + M_s^3)$, thus, for moderate-to-large $T$, we generally expect quadratic scaling with the number of inducing points and linear scaling with respect to the number of spatial and temporal locations \citep{hamelijnck2021spatio}. The random and quasi-random designs have trivial computational costs, and MES and IMSE computations take several minutes, even when using Markovian GPs and continuous optimization, as detailed in Appendix C \citep{waxman2025milsupplement}, rather than the typical discrete optimization with kernel-based GPs. We thus conclude that, while MIL is generally more expensive than other OED methods, its cost is on the same order of magnitude as that of other model-based approaches in our case study.

\subsection{Joint Learning is Important}

A primary motivation for our method was that greedy learning algorithms, such as BAD may provide strongly suboptimal results in the spatiotemporal setting, while our method instead provides a tractable approach to joint optimization.

\begin{figure}[t]
    \centering
    \begin{minipage}{0.48\linewidth}
    \centering \includegraphics[width=0.8\linewidth]{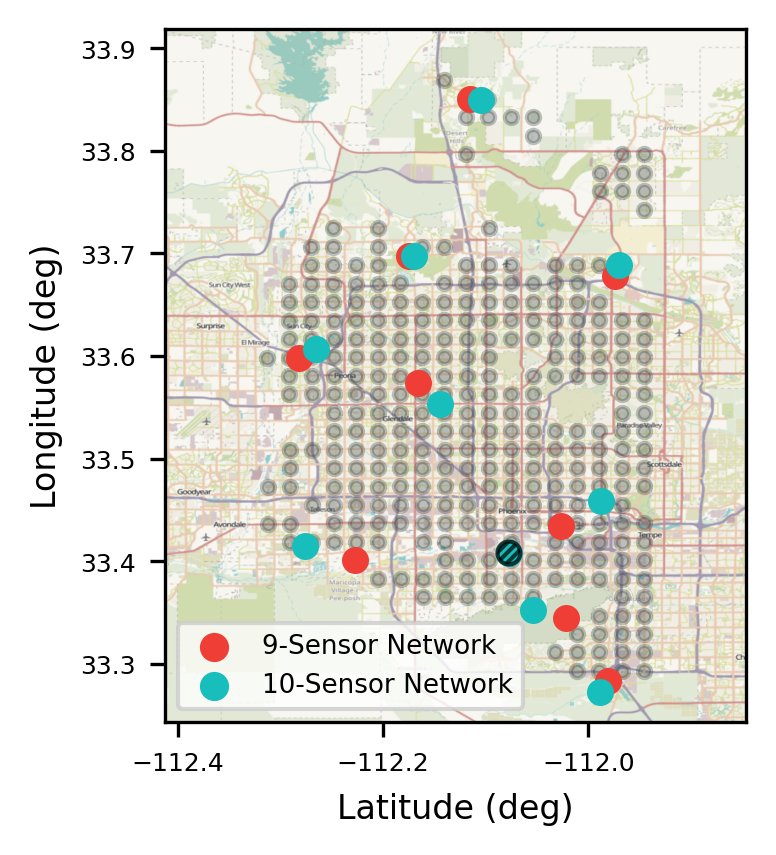}
    \captionof{figure}{Resulting inducing point locations when using 9 and 10 sensors. For visualization, we paired inducing point locations via the Hungarian algorithm, and highlighted the inducing point location least like the set of 9 inducing point locations.}
    \label{fig:plot_9_to_10}
    \end{minipage}%
    \hfill
    \begin{minipage}{0.48\linewidth}
        \centering
        \includegraphics[width=0.8\linewidth]{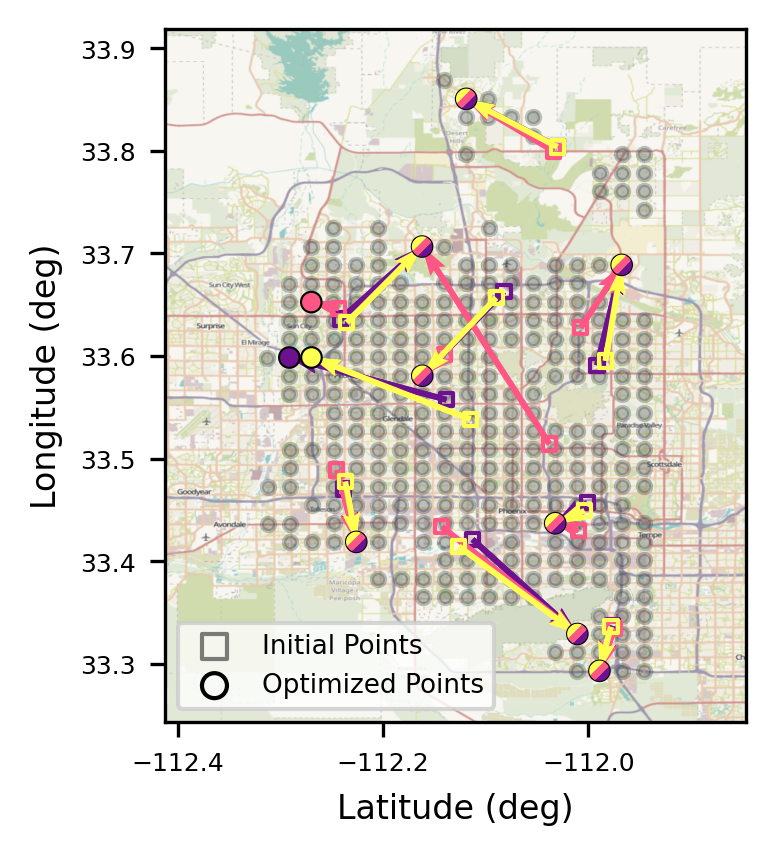}
        \captionof{figure}{The initial inducing point locations (squares), mapped to their corresponding values after optimization and discretization (circles) by arrows. Three different initializations were used, with colors denoting different random seeds.\vspace{1em}}
        \label{fig:maps_to_optimal}
    \end{minipage}
\end{figure}

\begin{figure}[t]
    \centering
    \begin{minipage}{0.49\linewidth}
    \centering \includegraphics[width=0.8\linewidth]{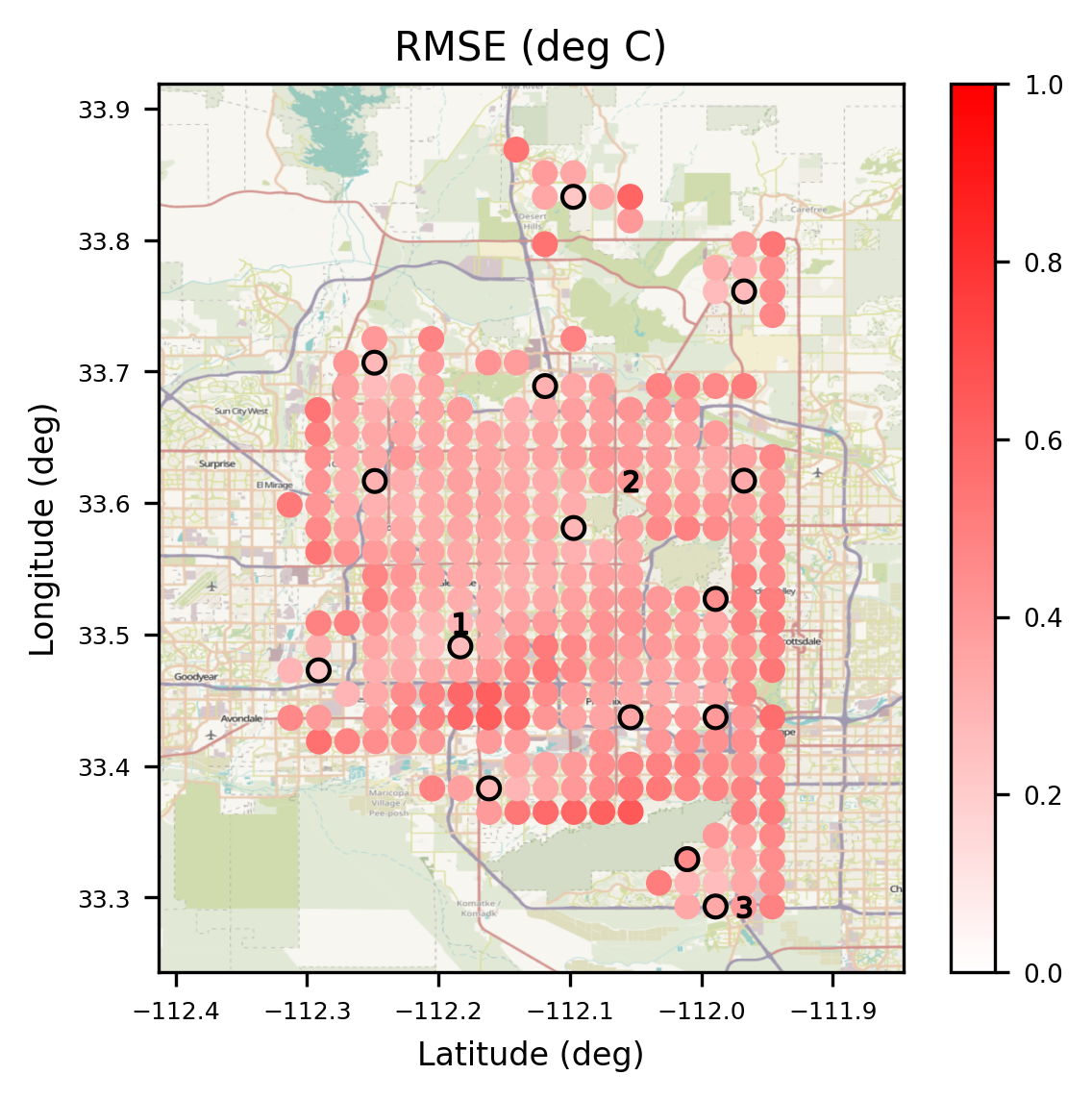}
    \captionof{figure}{The RMSE (averaged temporally across the entire test period) as a function of spatial location for an MIL-optimal configuration of $15$ sensors. Sensor locations are circled, and numbered labels ($1$, $2$, and $3$) indicate example locations for which temporal predictions are plotted in \cref{fig:example_timeseries}}
    \label{fig:spatial_rmse}
    \end{minipage}%
    \hfill
    \begin{minipage}{0.456\linewidth}
        \centering
        \includegraphics[width=0.8\linewidth]{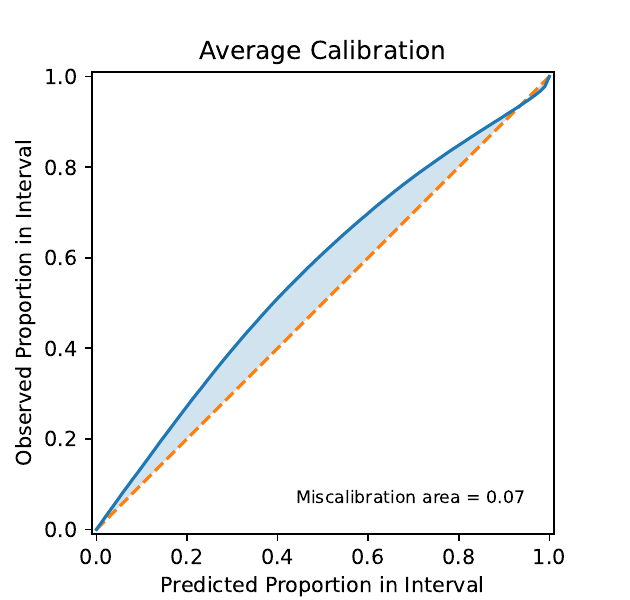}
        \vspace{2em}
        \captionof{figure}{A calibration plot for the predictions shown in \cref{fig:spatial_rmse}. The calibration plot shows the mismatch between the predicted proportion of data in a credible interval (x-axis) and the actual amount of data in the credible interval (y-axis).}
        \label{fig:miscalibration_plot}
    \end{minipage}
\end{figure}

To visually argue the importance of this joint learning, we optimized the inducing point locations for 9 and 10 sensors separately. The resulting inducing points, shown in \cref{fig:plot_9_to_10}, demonstrate how the optimal locations of inducing points may change when an additional inducing point is added. In other words, the first 9 optimal sensor locations in each network are not the same. 

\subsection{Visualizing the Spatiotemporal Performance}

Another way to quantify the performance of our approach is to examine the spatial error distribution. In \cref{fig:spatial_rmse}, we show the RMSE at each test location, averaged across the temporal dimension. Notably, locations with high errors are not necessarily those that are furthest from a sensor, underscoring the importance of model-based utilities. 

\begin{figure}[h]
    \centering
    \includegraphics[width=0.85\linewidth]{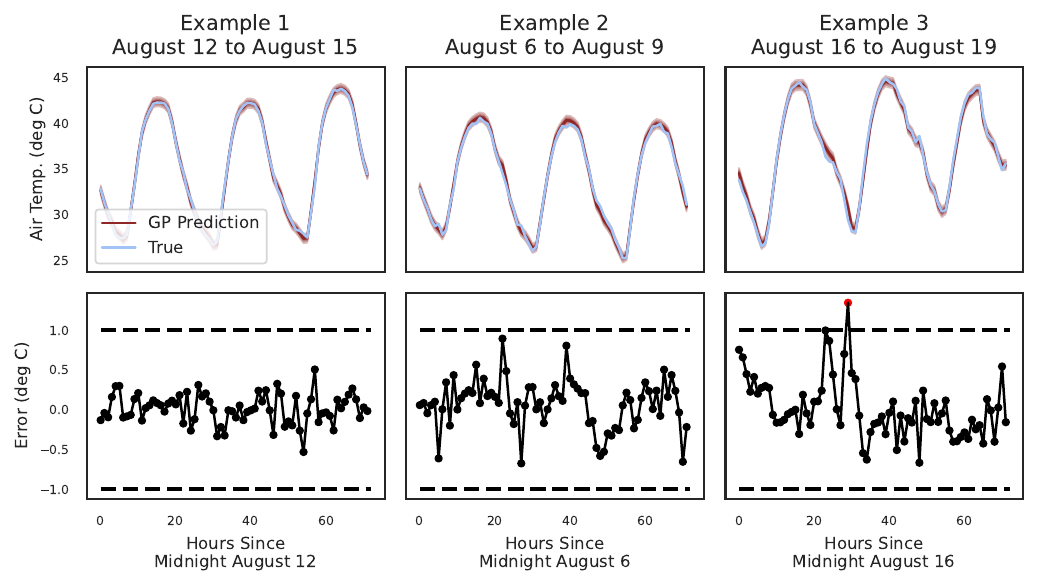}
    \caption{{\textbf{(Top)}} The GP predictions and true data over a randomly selected period of three days across three randomly selected locations (pictured in \cref{fig:spatial_rmse}) and {\textbf{(bottom)}} the corresponding errors in degrees Celsius.}
    \label{fig:example_timeseries}
\end{figure}

In \cref{fig:example_timeseries}, we visualize the predictions and errors for three randomly selected locations (labeled in \cref{fig:spatial_rmse} as $1$, $2$, and $3$) over three randomly selected days of the test period. We find strong predictive performance and well-calibrated uncertainty estimates. A calibration plot (\cref{fig:miscalibration_plot}) shows strong calibration, with a miscalibration area \citep{chung2021uncertainty} of $0.067$. From the calibration plot, we also see that our predictions tend to be underconfident in this case study, in the sense that our predictive variances are overly conservative.

\begin{figure}[t]
    \centering
    \includegraphics[width=0.24\linewidth]{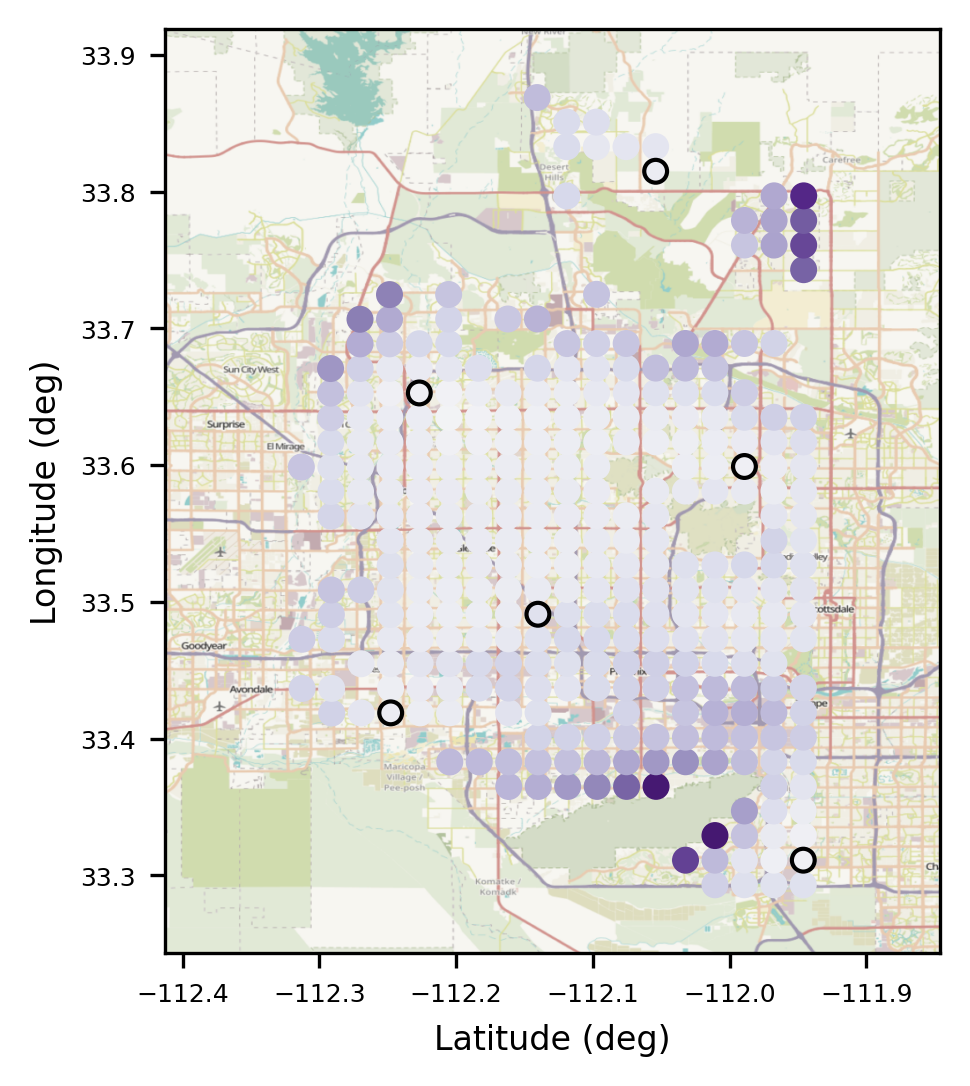} 
    \includegraphics[width=0.225\linewidth,trim={0.55cm 0 0 0},clip]{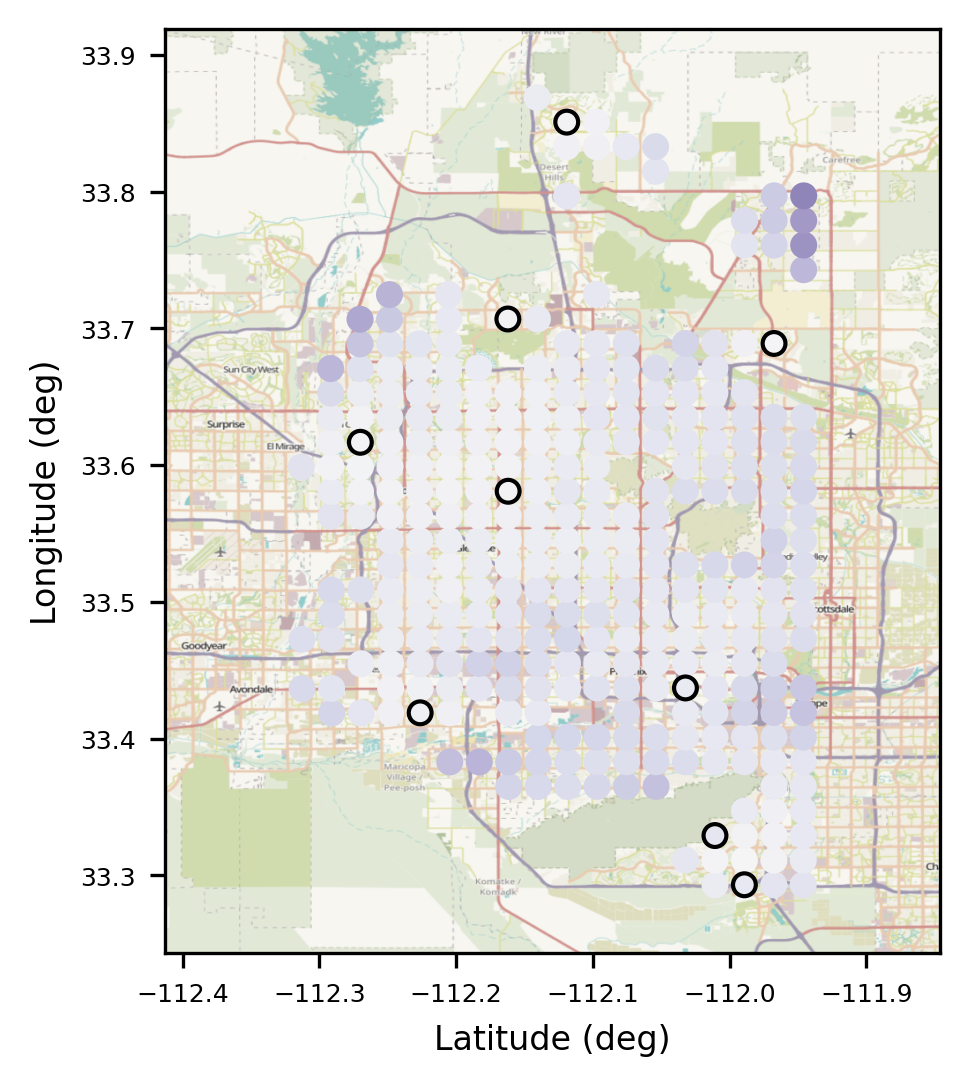} 
    \includegraphics[width=0.225\linewidth,trim={0.55cm 0 0 0},clip]{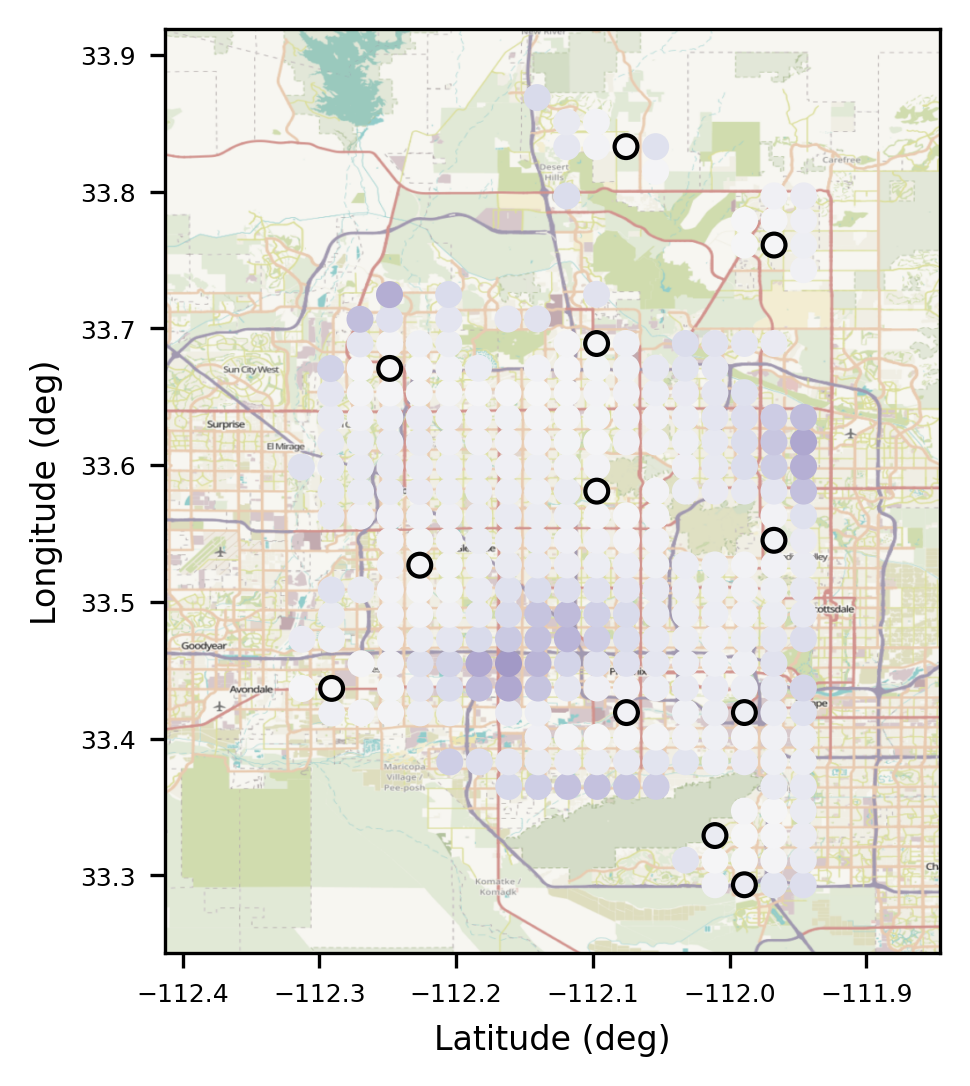} 
    \includegraphics[width=0.265\linewidth,trim={0.55cm 0 0 0},clip]{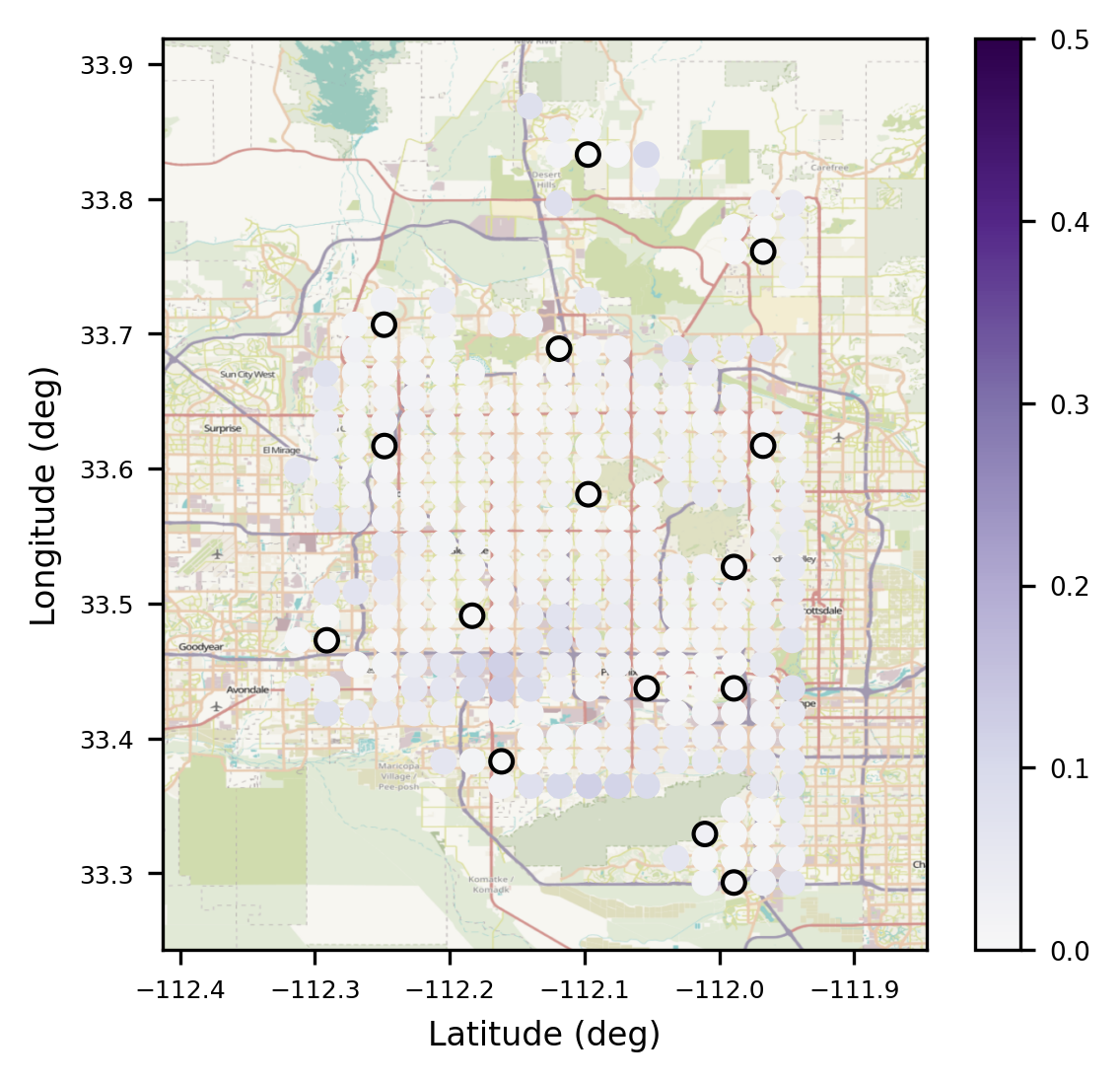} 
    \caption{The proportion of temporal instances which exhibit ``extreme errors'' (defined as greater than 1 degree Celsius) for sensor networks with 6, 9, 12, and 15 sensors.}
    \label{fig:extreme_errors}
\end{figure}

Also of interest to scientific stakeholders is the number of ``extreme'' errors in the predictions of a model. For example, maintaining subdegree accuracy, while ultimately arbitrary, may be of interest in validating a model. We visualize the number of errors greater than 1 degree Celsius as the number of sensors grows from 6 to 15 in \cref{fig:extreme_errors}. The results clearly show that increasing the number of sensor locations greatly reduces the number of extreme errors.

\section{Discussion \& Conclusions} \label{sec:discussion}
Recent trends in science and computation have led to new, challenging modalities for experimental design. For one such challenge, we have presented a novel approach to spatiotemporal experimental design that leverages available physics-based simulations as data to determine sensor placements. We further showed how this relates to the many existing approaches of Bayesian experimental design and how it may be information-theoretically justified as minimizing a type of ``information loss'' with respect to the physics-based simulation. We demonstrated the applicability of our approach to climate sciences using a case study of 2-m air temperature sensing in Phoenix, Arizona. For this case study, we applied our method to data from a physics-based climate model and showed that the deployment of a relatively small number of sensors at optimized locations can result in strong predictive performance.

Key to the scalability of our method is a Gauss-Markov prior for the spatiotemporal process, for which recent work has shown that efficient and tractable variational inference is possible when inducing points are spatially fixed. Using the sparse formulation given by GPs, we avoid the combinatorial optimization of sensor locations and instead solve a relaxed, continuous optimization problem amenable to gradient-based inference. 

The use of a Markovian GP has significant advantages as it  connects to a rich classical literature on BED, allowing for the incorporation of existing sensors and straightforward extensions to multi-fidelity data. One application of multi-fidelity approaches within our framework is using additional data from new deployments, where scientific data not belonging to the designed sensor network may become available. 
One advantage of our proposed method is that it directly provides a GP predictive model as a byproduct, which has been well studied as a surrogate model and offers established methods to mix real and simulated data. The most popular approach with GPs in this scenario is to model the error of the physics-based simulator itself with another GP, introduced by \citet{kennedy2001bayesian}. Since the sum of two GPs is itself a GP, this method retains the convenient predictive structure of our approach.

In our case study, we show that the resulting optimal sensor locations lead to significant advantages over existing approaches. We further show that, with respect to physics-based model data, the resulting surrogate model predictions are accurate and well-calibrated. This suggests that our method, combined with real observations from the optimized sensor network, may be used as a powerful tool for evaluating the performance of computationally expensive physics-based models, in the sense that large disagreements with field observations can be attributed to deficiencies in the physics employed. This is scientifically valuable, as it provides a cost-effective way to interrogate physics-based models and identify areas where their improvement is necessary.

While the methodological tools of Markovian GPs and variational inference are key to the development of our method, they also introduce technical constraints that are interesting topics for future work. For example, the ST-SVGP work of \citet{hamelijnck2021spatio} relies on the separability assumption of the kernel, which is unlikely to hold in many applications. One path for future research on this problem is to consider mixture-of-experts that draw from prior knowledge of temporal nonstationarity, e.g., by considering different models for the times when an urban area is ``heating'' or ``cooling.'' In a similar fashion, we may consider introducing more complex spatial kernels in future applications, e.g., through deep kernel learning \citep{wilson2016deep}.

\begin{acks}[Acknowledgments]
We would like to thank Francisco Salamanca Palou for providing the physics-based weather model output used for this analysis.
\end{acks}

\begin{funding}
The first and third authors of this work were supported by the Department of Energy Office of Science, Office of Biological and Environmental Research’s Urban Integrated Field Laboratories research activity, under field work proposal EE686EECP ``Southwest Urban Corridor Integrated Field Laboratory''. The second and fourth authors were supported by the National Science Foundation under Award 2212506.
\end{funding}

\begin{supplement}
\stitle{Appendices to ``Designing an Optimal Sensor Network via Minimizing Information Loss''}
\sdescription{The appendices comprise five parts: Appendix A includes further discussion of Bayesian optimal experimental design; Appendix B illustrates how additional sensors may be incorporated in our optimization scheme; Appendix C explicate the baselines used in our experiments; Appendix D presents preliminary experiments with deep kernel learning; and Appendix E show experiments quantifying the effects of simulator error, in line with our discussion in Section 3.3 of the main manuscript.}
\end{supplement}

\bibliographystyle{ba}
\bibliography{main}

\clearpage
\appendix
\section{Further Discussion of Bayesian Optimal Experimental Design} \label{app:more_boed}
In this appendix, we continue our discussion of utility-based sensor placement from Section 2.1 with several common examples.

For the sensor placement problem, one particularly common choice is to minimize the conditional entropy $H(\mathbf{Y}_\text{test} \given \mathbf{Y}_N)$ directly. Once again, if $\mathcal{X} = \mathbf{X}_n \sqcup \mathbf{X}_\text{test}$, we may decompose the differential conditional entropy as 
\begin{equation}\label{eq:entropy_decompose}
    H(\mathbf{Y}_\text{test} \given \mathbf{Y}_n) = H(\mathbf{Y}_\mathcal{X}) - H(\mathbf{Y}_{n}).
\end{equation}
If we assume that $H(\mathbf{Y}_\mathcal{X})$ is bounded and independent of our choice of $\mathbf{X}_n$, then minimizing $H(\mathbf{Y}_\text{test} \given \mathbf{Y}_n)$ 
is equivalent to maximizing $H(\mathbf{Y}_n)$. %
With the Gaussian assumption about ${\bf Y}_n$, minimizing the conditional entropy in \cref{eq:entropy_decompose} is equivalent to maximizing the determinant of the prior covariance matrix,
\begin{equation} \label{eq:MES_defn}
    \mathbf{X}^\text{opt}_n = \argmax_{\mathbf{X}_n \subset \mathcal{X}} \{\logdet \mathbf{K}_\text{prior}(\mathbf{X}_n)\}.
\end{equation}
This criterion, known as maximum entropy sampling (MES), notably does not depend on the observed data explicitly \citep{shewry1987maximum,sebastiani2000maximum}. This is computationally convenient but philosophically puzzling, since the point of a statistical approach in the first place is to provide observation-driven sensor locations rather than geometrically heuristic sensor locations. 

\begin{remark}
    \citet{shewry1987maximum,sebastiani2000maximum} motivate MES as being equivalent to EIG if the entropy $H(\Ytest)$ is independent of the design parameters. However, when applied to the sensor design problem, $H(\Ytest)$ clearly depends on the sensor locations $\mathbf{X}_n$, since $\mathbf{X}_\text{test} \triangleq \mathcal{X} \setminus \mathbf{X}_n$. Thus, we cannot claim that MES is equivalent to maximizing the EIG.
\end{remark}

\begin{remark}
    If we allow for heterogeneous additive noise, i.e., $\widetilde{Y}(\mathbf{x}) = Y(\mathbf{x}) + \varepsilon(\mathbf{x})$, the decomposition above cannot be applied directly, as the joint entropy of $(\widetilde{{\bf Y}}_n, \mathbf{Y}_\text{test})$ now depends on the choice of $\mathbf{X}_n$. We will address this by proposing an optimization method over the random field $Y(\cdot)$ which directly incorporates the remaining entropy of noisy observations.
\end{remark}

In the Gaussian case, maximum entropy sampling (MES) criterion can be interpreted with the utility function,
\begin{equation*}
    U_{\text{MES}}(\mathbf{X}_n) = \logdet {\bf K}_\text{prior}(\mathbf{X}_n).
\end{equation*}
The classical Bayesian D-optimality \citep{chaloner1995Bayesian} admits a prediction-space analogue in the GP setting,
\begin{equation*}
    U_{\text{D}}(\mathbf{X}_n) = \logdet {\bf K}_\text{post}(\mathbf{X}_\text{test}).
\end{equation*}
Finally, the Integrated Mean Squared Error (IMSE)  instead uses the utility function
\begin{equation*}
    U_{\text{IMSE}}(\mathbf{X}_n, \Yn) = -\int_{{\bf X}_\text{test}} \mbox{var}({Y}(\mathbf{x}) \given \mathbf{Y}_n) \, d\mathbf{x},
\end{equation*}
where the integral is understood to be with respect to the counting measure for discrete choices of $\mathbf{X}_\text{test}$ (i.e., the corresponding sum). In practice, if $\mathbf{X}_\text{test}$ is continuous, this integral must be approximated via quadrature over a grid of points $\mathbf{X}_\text{grid}$, where the integral becomes the trace of the posterior covariance matrix, i.e., 
\begin{equation*}
    U_{\text{IMSE}}(\mathbf{X}_n, \Yn) \approx -\trace\left[ \mathbf{K}_{\text{post}}(\mathbf{X}_\text{grid})\right].
\end{equation*}

\section{Illustrating The Incorporation of Existing Sensors} \label{app:include_existing}

\begin{figure}[tbp] %
    \centering
    \begin{subfigure}{0.35\textwidth} %
        \centering %
        \includegraphics[width=\linewidth]{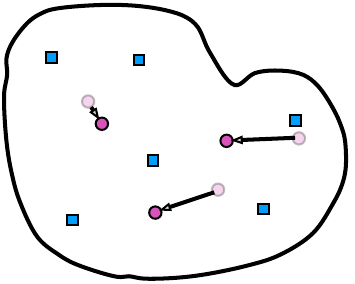}
        \vspace{0.5em}
    \end{subfigure}%
    \hspace{0.1\linewidth}
    \begin{subfigure}{0.35\textwidth} %
        \centering %
        \includegraphics[width=0.9\linewidth]{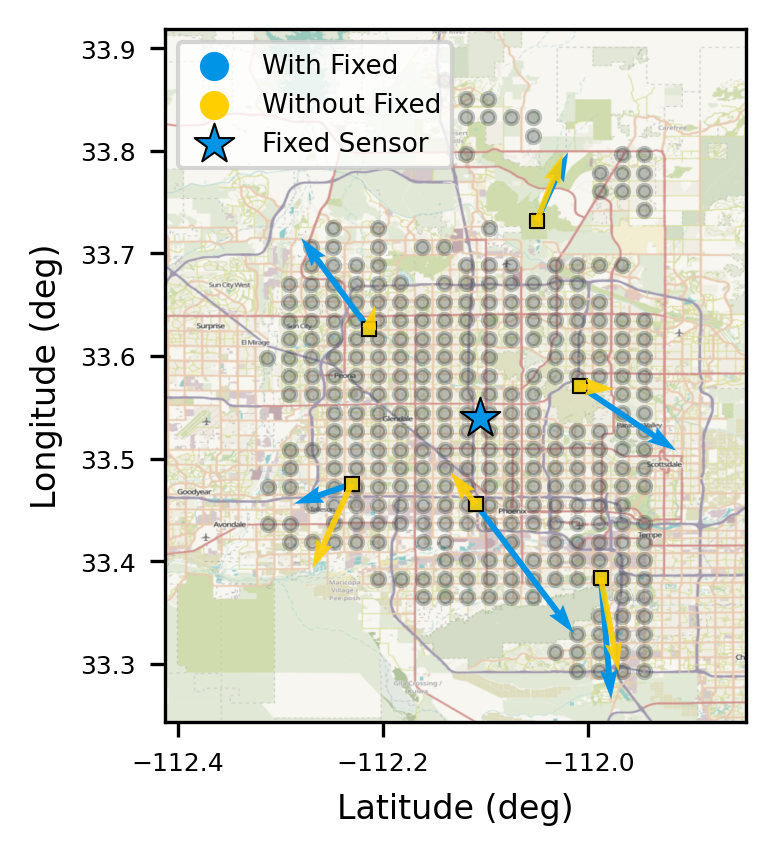}
    \end{subfigure}
    \caption{{\textbf{(Left)}} A sketch of how existing observation locations can be incorporated in our optimization scheme. New locations (pink circles) are allowed to move during optimization, but existing locations (blue squares) are fixed during the entire optimization period. {\textbf{(Right)}} A real example corresponding to the case study in Section 6: sensor designs including a central fixed sensor (blue) clearly avoid the center of the domain more than sensor designs trained with no existing sensors present (yellow).}
    \label{fig:existing_locations_sketch}
\end{figure}

In this appendix, we illustrate the incorporation of existing sensor locations into the MIL framework. In particular, in \cref{fig:existing_locations_sketch}, we sketch the evolution of sensor locations with the existing locations present. We also include a real example based on our case study that shows the effect on learned sensor networks with and without an existing sensor placed in the center of the domain, reinforcing effectiveness.

\section{Baselines for Case Study} \label{app:baselines}

We compare two popular strategies in spatial OED, also employed in the spatio-temporal setting \citep{mateu2012spatio}, namely (a) maximizing the entropy of the distribution of the observations at the design \citep{jin2012entropy}, and (b) minimizing the variance of the predictions given the design \citep{heuvelink2012sampling}. We refer to the former as MES and to the latter as IMSE. Certainly, these model-based criteria depend on the chosen model. To keep comparisons fair, we choose the same spatio-temporal GP prior used in MIL (i.e., the ST-SVGP \citep{hamelijnck2021spatio}, using the same kernels as in the main manuscript). Furthermore, these two strategies require an initial set of real observations to estimate model hyperparameters and compute a predictive distribution. This is fundamentally different from our approach, where we assume we are in a {\it pre-observation} setting and use synthetic data. Hence, for a fair comparison in MES and IMSE, we should use real data both to find the optimal design and to evaluate its effectiveness. On the other hand, in MIL, we should use only synthetic data and then evaluate it against real observations. Since we are using a single synthetic dataset, the results of this experiment slightly favor our MIL.

In the following, we provide details on the implementation of MES and IMSE. As mentioned above, we use the same GP prior as in all other experiments.

\subsection{Maximum Entropy Sampling}
In MES, we initialize the model by fitting it to a dataset containing temperatures at $N_\text{init}=5$ spatial locations during the first month (June). These initial points were chosen in a space-filling fashion by running the k-means algorithm on the entire spatial grid and selecting the grid points closer to each centroid \citep{jin2017k}. Then, the predictive distribution (which can also be seen as a type of conditional prior) over the next month (July) is used to select the design. More specifically, we choose $N_\text{add}\in\{1,4,7,10\}$ such that the log-determinant of the posterior covariance matrix at the input design is maximized. 
The total number of sensors $N_\text{tot} = N_\text{init} + N_\text{add}$ is $N_\text{tot}\in\{6,9,12,15\}$.

To ensure computational traceability, we solve the MES in a continuous space. One well-known drawback of MES is its proclivity to push points to the boundary of the design space -- to create a motivated design space, we compute the convex hull over all spatial locations and use projected gradient descent \citep[Ch. 2]{bertsekas1999nonlinear} for tractability. 

Due to the separability assumption on the kernel and the prediction set being a spatiotemporal grid, the full kernel may be symbolically constructed as a Kronecker product, $K = \Sigma_S \otimes K_T$, for some covariance matrix $\Sigma_S$ involving only spatial elements \citep[Eq. (27)]{hamelijnck2021spatio}. The log-determinant thus decomposes as $\log\det K = N_T \log\det \Sigma_S + N_S \log\det K_T$ \citep[Sec. 2]{minka2000old}; since the second part of this sum does not depend on the spatial design, we may optimize purely with respect to $\log\det \Sigma_S$. This allows for computational tractability, as it does not materialize a large covariance matrix or compute its determinant. It further highlights a drawback of the MES approach, namely that it does not capture any temporal information in its spatial designs.

\subsection{Integrated Minimum Square Error}

In IMSE, we proceed similarly and fit the model to obtain the predictive distribution over the next month (July). The design is chosen such that the trace of the posterior covariance matrix, augmented with ``imagined'' observations from July at a set of predefined locations, is minimized. Notably, because of the structure of GP predictive variances, it does not matter what these imagined observations are, so long as the kernel is fixed. This allows us to compute the IMSE-optimal design exactly with Markovian Gaussian processes. This, in turn, provides $\mathcal{O}(N_t)$ scaling over the design, which offers a computational advantage over using a conventional GP in spatiotemporal design. This is, to the best of our knowledge, a methodologically novel improvement. 

Note that for computing this criterion, we need to choose a fixed prediction grid to sum over, conditional on the chosen design. We chose the entire spatial grid as the prediction grid (314 points). This aligns with the evaluation criteria (both RMSE and NPLL), which are computed over the same grid.

\section{Preliminary Experiments With Deep Kernels}
\label{app:dkl}
In Section 6, we included a case study using a separable kernel, with a mixture Hida-Mat\'ern kernel as the temporal sub-kernel and a Mat\'ern kernel for the spatial sub-kernel. While separability and a stationary temporal kernel are key assumptions we make for convenient variational algorithms, it was noted that the spatial kernel need not be stationary. Indeed, the spatial kernel can be an arbitrary positive semi-definite kernel, e.g., as learned via deep kernel learning \citep{wilson2016deep}.

To illustrate this, we performed preliminary experiments using a deep kernel for the spatial kernel. This allows for extremely flexible spatial relationships by adopting a base kernel $\kappa_0(\cdot, \cdot)$ to use inputs warped by a neural network $\mathrm{NN}_\theta$, i.e.,
\begin{equation}
    \kappa(\V{x}, \V{x}') = \kappa_0\left(\mathrm{NN}_\theta(\V{x}), \mathrm{NN}_\theta(\V{x}')\right).
\end{equation}

In our proof-of-concept, we use a Mat\'ern-3/2 kernel for $\kappa_0$, and a feedforward neural network using layer widths of $64$, $32$, and $16$ for $\mathrm{NN}_\theta$, with leaky ReLU activation functions. 

\begin{figure}[t]
\centering
\begin{minipage}{0.48\linewidth}
    \centering
    \includegraphics[width=\linewidth]{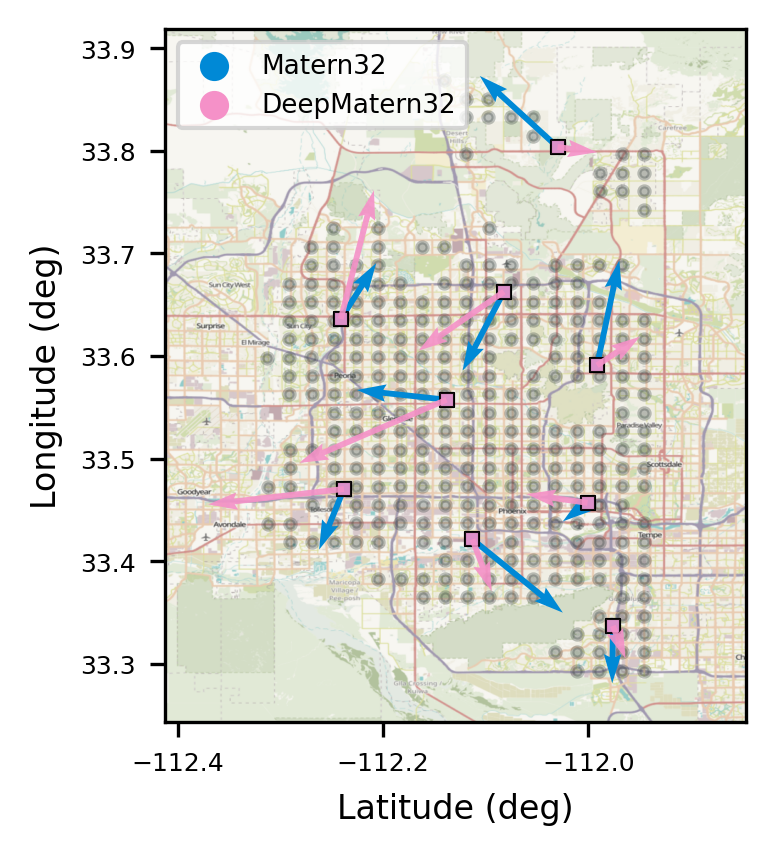}
    \caption{A comparison in MIL-optimal sensor locations between a stationary and non-stationary kernel.}
    \label{fig:optimal_pts_dkl}
\end{minipage}\hfill
\begin{minipage}{0.48\linewidth}
    \centering
    \begin{tabular}{lcc}
    \toprule
    Model & RMSE & NLPD \\
    \midrule
    Mat\'ern-3/2      & 0.5921 & -0.9468 \\
    Deep Mat\'ern-3/2 & 0.6049 & -0.8850 \\
    \bottomrule
    \end{tabular}
    \caption{Performance comparison between a stationary Mat\'ern-3/2 kernel, and deep Mat\'ern-3/2 kernel.}
    \label{tab:metrics_dkl}
\end{minipage}
\end{figure}

\begin{figure}
    \centering
    \includegraphics[width=0.95\linewidth]{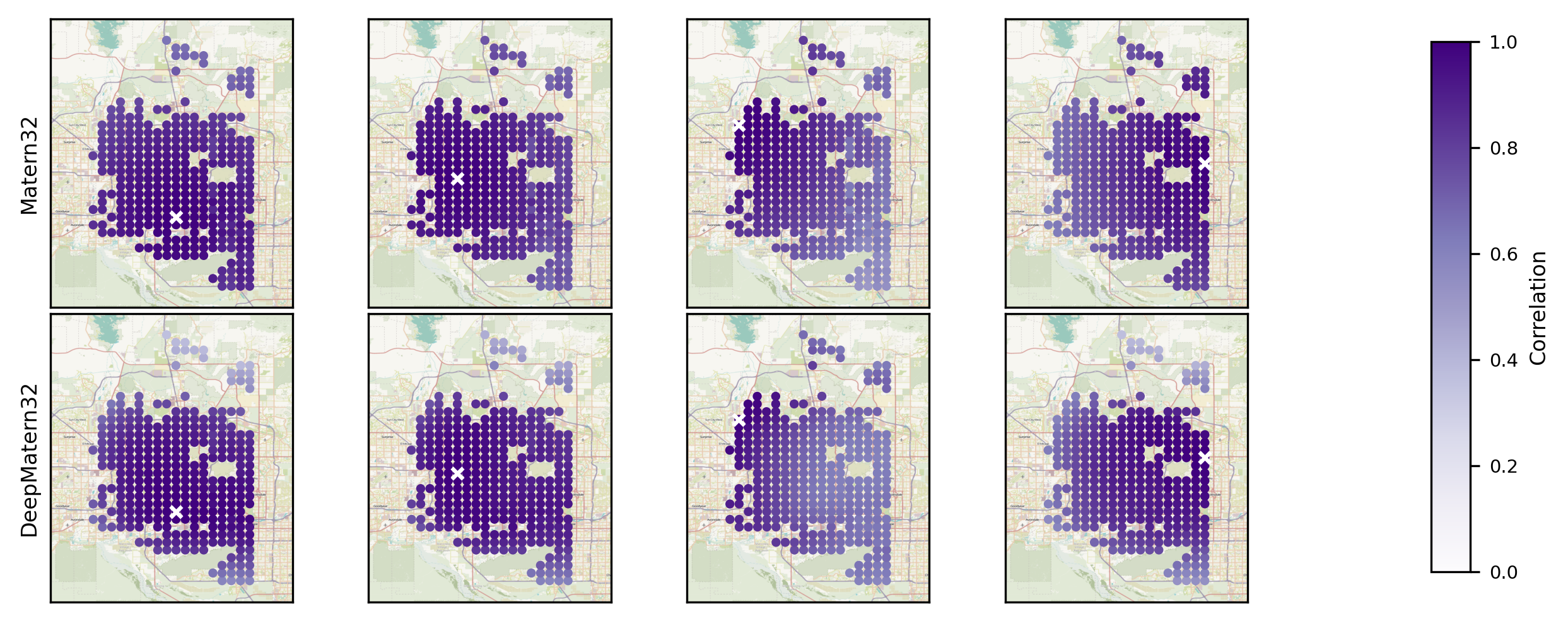}
    \caption{A comparison of the learned Mat\'ern-3/2 and Deep Mat\'ern-3/2 spatial kernels. The Deep Mat\'ern-3/2 exhibits some mild non-stationarity, while the Mat\'ern-3/2 kernel does not.}
    \label{fig:ns_kernel}
\end{figure}

We repeat the experiment in Section 6, training using a Mat\'ern-3/2 kernel and a deep Mat\'ern-3/2 kernel for $250$ epochs, using a reducing learning rate schedule for the deep Mat\'ern-3/2 kernel. We obtain visually distinct optimal experimental designs (\cref{fig:optimal_pts_dkl}), with slightly degraded performance (\cref{tab:metrics_dkl}). We visualize the spatial non-stationarity of the resulting kernel in \cref{fig:ns_kernel}. In our experiments, we find that deep kernels are, somewhat unsurprisingly, more sensitive to hyperparameters and less stable than their stationary counterparts.

Overall, this proof-of-concept experiment illustrates that deep kernels show promise in learning non-stationary spatial covariances and alternative MIL-optimal configurations, but it requires further study to better understand the decisions surrounding stable training and kernel design.

\clearpage
{

\section{Experiments with ``Simulator Error''} \label{app:simulator_error}

To further investigate the robustness of the MIL selection criterion under simulator misspecification, we conducted a controlled numerical study. Specifically, we constructed a toy example based on our Phoenix, Arizona case study by introducing GP-based simulator error, i.e., $f_\text{obs}(\V{x}, t) = f_\text{true}(\V{x}, t) + f_\text{err}(\V{x}, t)$. This is in line with classical assumptions of simulator error \citep{kennedy2001bayesian}, with noise in our case being drawn from a separable Mat\'ern-3/2 kernel,
\begin{align}
    \kappa(\mathbf{x}, t, \mathbf{x}', t') &= \kappa_s(\mathbf{x}, \mathbf{x}') \kappa_t(t, t') \\
    &= \sigma_\text{noise}^2 \kappa_\text{Mat-3/2}(\V{x}, \V{x}'; \ell_{s}) \kappa_\text{Mat-3/2}(t, t'; \ell_{t}). 
\end{align}
To evaluate the impact of misspecification, we compared the design points selected by MIL across different simulator realizations. To better measure how simulator error directly influences performance, we chose the training set to be exactly the evaluation set, with the added GP noise. The noise uses spatial lengthscales -- in normalized units -- of $\ell_s \in \{0.1, 1.0\}$, representing localized and spatially diffuse noise, respectively, and temporal noise in $\ell_t \in \{1.0, 36.0\}.$ The training set consists of the first 7 days of June 2013. We evaluate combinations of small and large noise variances with localized and globalized spatial and temporal lengthscales. Each experiment deploys $9$ sensors.

We extract a set of ``golden'' hyperparameters by first fitting an ST-SVGP to the ground-truth data to isolate predictive error stemming from hyperparameter learning and predictive error from inferior sensor networks. We repeat this process for 10 replications, which we note provide legitimately different noise characteristics each time. 

Results, in terms of RMSE and NLPD, are shown in \cref{fig:sim_error_exp_results}. The results support our theoretical arguments, i.e., that moderate simulator error results in acceptable error rates, still comparable to baseline methods, and that localized noise and high signal-to-noise ratios are both crucial. We additionally visualize the sensor locations learned in each noise setting for a specific realization in \cref{fig:sim_error_locations}. We can easily verify that the noise variance has legitimate effects on the resulting sensor locations.

\begin{figure}
    \centering
    \includegraphics[width=0.95\linewidth]{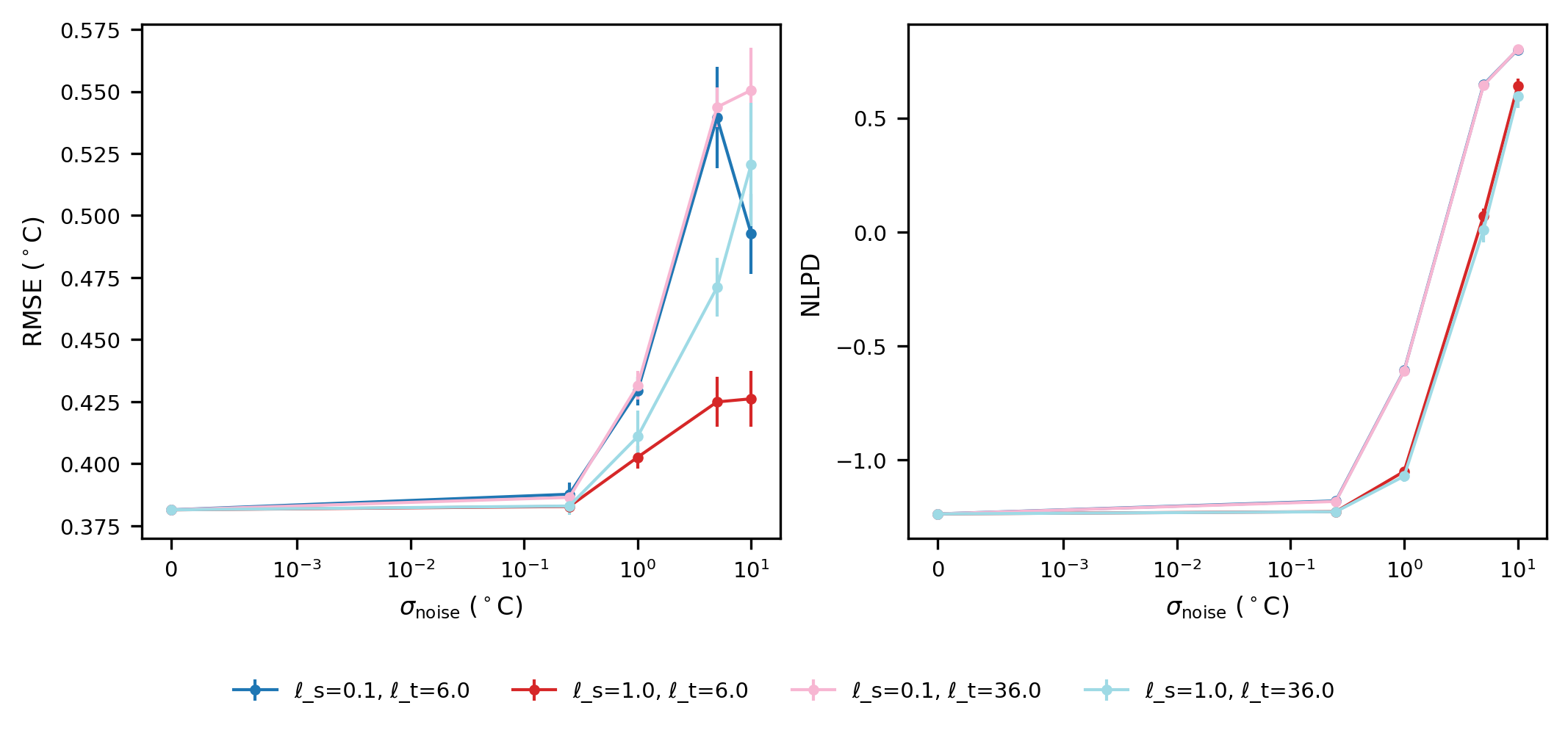}
    \caption{Results of the ``simulator error'' ablation study. Error bars denote one standard deviation over 10 realizations.}
    \label{fig:sim_error_exp_results}
\end{figure}

\begin{figure}
    \centering
    \includegraphics[width=0.55\linewidth]{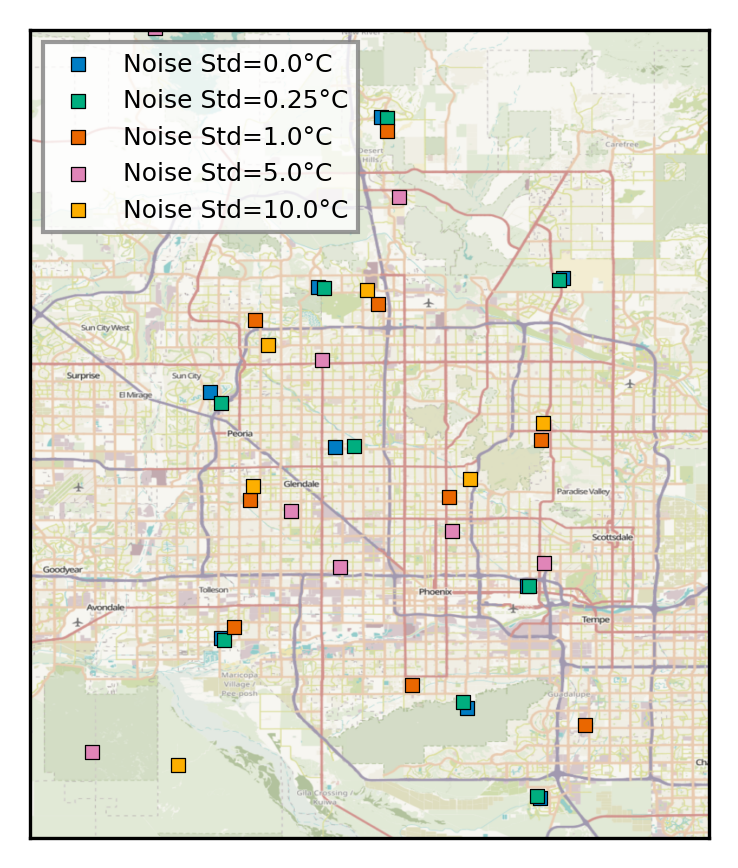}
    \caption{Sensor locations under various noise variance settings, with $\ell_s = 1.0$ and $\ell_t = 36.0$.}
    \label{fig:sim_error_locations}
\end{figure}

}

\end{document}